\documentclass[prb,superscriptaddress,reprint,longbibliography,nobibnotes]{revtex4-1} 

\usepackage{setspace} 
\usepackage{graphicx}
\usepackage{amsmath}
\usepackage{color}
\usepackage{amsmath}
\usepackage{amssymb}
\usepackage{verbatim}
\usepackage{latexsym}
\usepackage{braket}
\usepackage{enumerate} 
\usepackage{bm} 
\usepackage{ulem} 
\usepackage[dvipsnames]{xcolor}

\graphicspath{{./img/}}

\begin{document}

\title{
Altermagnetic Shastry-Sutherland fullerene networks
}

\newcommand{\TCM}{Theory of Condensed Matter Group, Cavendish Laboratory, University~of~Cambridge, J.\,J.\,Thomson Avenue, Cambridge CB3 0HE, UK}

\newcommand{\Pet}{Peterhouse, University of Cambridge, Trumpington Street, Cambridge CB2 1RD, UK}


\author{Jiaqi Wu}
\thanks{These authors contributed equally.}
\affiliation{\Pet}

\author{Alaric Sanders} 
\thanks{These authors contributed equally.}
\affiliation{\TCM}

\author{Rundong Yuan} 
\affiliation{\TCM}

\author{Bo Peng}
\email{bp432@cam.ac.uk}
\affiliation{\TCM}

\date{\today}

\begin{abstract}
The interplay between quantum magnetism and many-body physics is of fundamental importance in condensed matter physics. 
Molecular building blocks provide a versatile platform for exploring the exotic quantum phases arising from complex orderings in frustrated lattices. Here we demonstrate a showcase system based on altermagnetic Shastry-Sutherland fullerene networks, which can be constructed from a C$_{40}$ molecular synthon with two effective spin-1/2 sites due to the resonance structures. The charge-neutral, pure-carbon systems exhibit an altermagnetic ground state with fully compensated spins arranged in alternating C$_{40}$ units in a 2D rutile-like lattice, leading to $d$-wave splitting of the spin-polarised electronic band structure and strong chiral-split magnons. We report a rich phase diagram 
including altermagentic, quantum spin liquid, plaquette, and dimer phases, which can be accessed via moderate strains. Our findings open a new avenue for exploring quantum many-body physics based on scalable, chemically-feasible, molecular quantum materials.
\end{abstract}


\maketitle



Magnetism in low-dimensional systems provides fertile ground for the discovery of exotic quantum phases\,\cite{Sun2025,Su2025,Fu2025,Peng2025b,Song2025}. Beyond conventional ferromagnetism and antiferromagnetism, altermagnetism has emerged as a newly-identified class of collinear magnetism with compensated spin sublattices but finite momentum-dependent spin polarisation\,\cite{Hayami2019,Yuan2020,Smejkal2020,Ma2021a,Mazin2021,Simejkal2022,Simejkal2022a,Fedchenko2024,Krempasky2024}. The vanishing net magnetisation and symmetry-protected spin splitting offer tantalising opportunities for next-generation quantum devices with robust spin-polarised transport. Moreover, the interplay between altermagnetism and many-body physics leads to a rich variety of quantum phenomena such as superconductivity\,\cite{Bose2024,Pupim2025}, spin liquid\,\cite{Sobral2025}, Hubbard model\,\cite{Sato2024,Das2024,Giuli2025,He2025}, and Kondo lattice\,\cite{Zhao2025}. In this context, quantum magnets in the Shastry-Sutherland model\,\cite{SriramShastry1981,Miyahara1999,Knetter2000,Koga2000,Laeuchli2002,Dorier2008,Brassington2024} represent a showcase model to realise altermagnetism, as their neighbouring sublattices are connected by rotations and glide reflections in the absence of neither translational nor inversion symmetry\,\cite{Ferrari2024}. Furthermore, the Shastry-Sutherland lattice provides an ideal platform to explore the exotic quantum physics and complex many-body effects in altermagnets such as superconductivity\,\cite{Chung2004,Yang2008b}, strong correlation\,\cite{Liu2007}, quantum phase transition\,\cite{Lee2019,Shi2022}, and spin frustrations\,\cite{Pula2024,Corboz2025}.
Yet, realising altermagnetism in the Shastry-Sutherland lattice remains a challenge.

Carbon-based materials provide unpaired $\pi$ electrons in delocalised $p_z$ orbitals, which are typically found in defects such as edges, vacancies, or heteroatom sites that disrupt homogeneous electron density\,\cite{Yazyev2010,Slota2018,Ma2025}. However, it is challenging to synthesise these systems due to atomically-precise, periodic assembly of the defected carbon fragments into the $\pi$-conjugated frameworks. Comparing to carbon fragments, fullerene building blocks are promising alternatives due to their stable units and highly-tuneable structures. We have recently discovered approaches to systematically introduce magnetism into pure-carbon, charge neutral, fullerene monolayers by enforcing crystalline symmetry based on molecular-orbital theory\,\cite{Wu2025a} or by creating $\pi$ magnetism from resonating valence bond\,\cite{Peng2025d}, opening new avenues for realising exotic quantum phases such as quantum anomalous Hall effects in otherwise non-magnetic monolayers\,\cite{Pingen2025}. Most importantly, the inherent ability of assembling these molecules into periodic frameworks has been confirmed since the synthesis of polymeric C$_{60}$ monolayers\,\cite{Hou2022}, as demonstrated by their rich structural phases as predicted theoretically\,\cite{Peng2022c,Peng2023,Jones2023,Wu2025,Shearsby2025,Kayley2025,Shaikh2025,Peng2025c,Peng2025a} and synthesised experimentally\,\cite{Meirzadeh2023,Wang2023,Zhang2025}. Therefore, it would be insightful to introduce altermagnetism into fullerene-based networks as a promising platform to realise exotic models such as the Shastry-Sutherland lattice.

Here, we use fullerene building blocks to design quantum altermagnets in 2D Shastry-Sutherland lattice.
The key to the realisation of altermagnetism lies in the elegant design of a molecular network that incorporates both a C$_{40}$ unit, with unpaired electrons for $\pi$ magnetism, and a rutile-like 2D lattice, for constructing two sublattices with opposite spins. We show that the resonance structure of the C$_{40}$ cage provides six carbon sites for two unpaired electrons on the opposite sides of the molecule and stabilises two effective spin-1/2 sites, which, upon assembly into a rutile-like monolayer, realise an altermagnetic ground state. Our first-principles calculations reveal $d$-wave splitting in the electronic bands and strong chiral-split magnon dispersions. Most remarkably, the altermagnetic Shastry-Sutherland networks can be continuously tuned into the frustrated quantum spin liquid phase through moderate strain. Our work opens a new frontier that unites the chemistry of carbon-based molecules with the physics of strongly-frustrated quantum magnetism.


\begin{figure*}
\centering
\includegraphics[width=\linewidth]{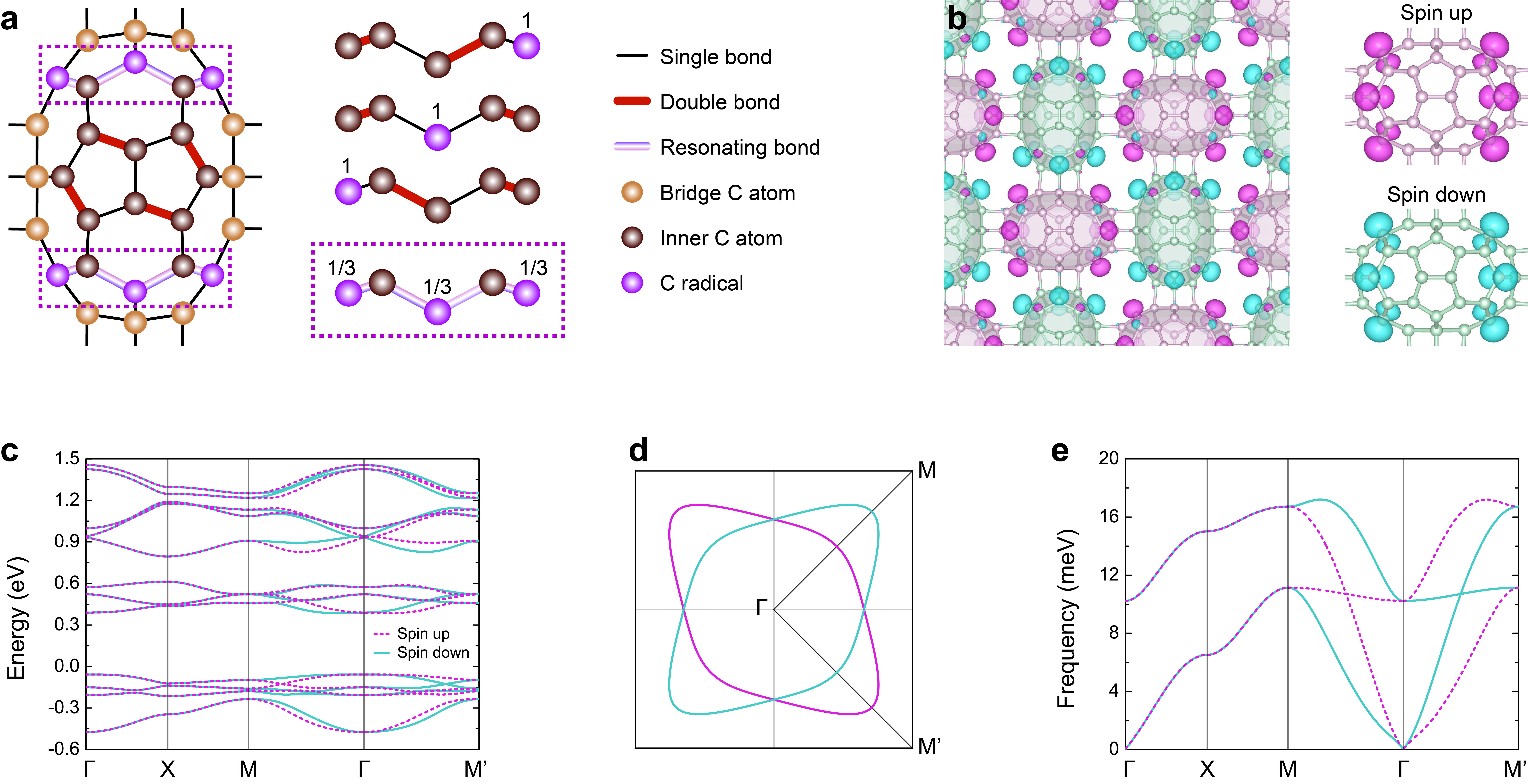}
\caption{
\textbf{Altermagnetism in monolayer C$_{40}$ fullerene networks.}
\textbf{a} C$_{40}$ building block and its resonance structures, \textbf{b} spin densities, \textbf{c} band structures, \textbf{d} iso-energy surface at 0.025\,eV below the VBM for $d$-wave order, and \textbf{e} chiral-split magnons. 
}
\label{crystals} 
\end{figure*}


\vspace{2mm}

\noindent\textbf{Altermagnetic fullerene networks.} Figure\,\ref{crystals}\textbf{a} shows a C$_{40}$ fullerene molecular synthon, which serves as a stable building block that has been synthesised decades ago\cite{Tomilin2001,Enyashin2008,Kharlamov2013}. Among the thirty carbon atoms on the top surface of C$_{40}$, there are twelve fully-saturated carbon atoms with $sp^3$ hybridisation for intermolecular bonds (marked in yellow). For the eight carbon atoms from the two five-membered rings in the centre of C$_{40}$, each atom is surrounded by three single bonds due to the $sp^2$ hybridisation, leaving eight $p_z$ electrons to form four double bonds as marked by red in Fig.\,\ref{crystals}\textbf{a}. Consequently, all twenty carbon atoms above are non-magnetic.

We then focus on the ten carbon atoms in two W-shaped chains, as highlighted in purple boxes in Fig.\,\ref{crystals}\textbf{a}. For each W-shaped cluster, there are three pairing schemes for two double bonds between the five carbon atoms, leaving one unpaired electron at the leftmost, middle and rightmost sites of the chain, respectively. The resonance structure leads to an averaged distribution of 1/3 $\mu_{\rm B}$ magnetic moment per purple carbon atom and a quantised total magnetic moment of 1 $\mu_{\rm B}$ for the group of five atoms in the W-shaped chain. Therefore, a C$_{40}$ unit has six magnetic carbon atoms with a total magnetic moment of 2 $\mu_{\rm B}$, as confirmed by first-principles calculations.

Afterwards, we construct closely-packed 
fullerene networks by rotating neighbouring C$_{40}$ building units by $\pi/2$, and such $C_4$ rotation creates a 2D rutile-like lattice similar to RuO$_2$\,\cite{Fedchenko2024}. The space-efficient arrangement of fullerene molecules is expected to stabilise the structure as found experimentally\,\cite{Hou2022,Meirzadeh2023}. Most interestingly, the closely-packed lattice leads to an altermagnetic ground state that is energetically favourable than the ferromagnetic, non-magnetic and other antiferromagnetic order, respectively. This suggests stronger intramolecular exchange interactions than the intermolecular coupling (as discussed later). Figure\,\ref{crystals}\textbf{b} shows the corresponding spin densities where magenta (cyan) iso-surface represents spin-up (spin-down) states. Within each C$_{40}$ unit, the evenly distributed spin densities among the six magnetic carbon atoms confirm the resonance structure of 1/3 $\mu_{\rm B}$ per magnetic carbon atom. The neighbouring C$_{40}$ units are connected by $C_4$ rotations and $G_x/G_y$ glide reflections in the absence of neither translational nor inversion symmetry, leading to compensated antiparallel magnetic order. Such an alternating order of the magnetic moments results in a zero net magnetisation. 
Additionally, for fullerene units with light carbon elements, the spin-orbit coupling approaches the non-relativistic limit. Therefore, the vanishing net magnetisation has a strong non-relativistic origin.

We next examine the band structures of altermagnetic C$_{40}$ networks in Fig.\,\ref{crystals}\textbf{c}. The spin-up and spin-down bands are doubly degenerate along the $\Gamma$--X--M high-symmetry paths. Along M--$\Gamma$--M', however, the symmetry-protected collinear compensated magnetic order in real space generates an alternating spin polarisation in the band structure in the reciprocal momentum space, and the spin-up and spin-down bands become split, breaking the time-reversal symmetry without magnetisation. We further compute the iso-energy surface at 0.025\,eV below the valence band maximum (VBM), which corresponds to moderate hole doping. As shown in Fig.\,\ref{crystals}\textbf{d}, the spin orientations at time-reversed, opposite momenta on the iso-energy surface are the same (non-time-reversed), while the spin-up and spin-down bands are symmetric, exhibiting typical features for $d$-wave order similar to RuO$_2$\,\cite{Fedchenko2024}. 

To further confirm the altermagnetism, we compute the magnon dispersion. As shown in Fig.\,\ref{crystals}\textbf{e}, the low-frequency magnons near $\Gamma$ show linear dispersion, which is a typical antiferromagnetic/altermagnetic feature\cite{Kittel1976}. 
The degenerate magnon bands along $\Gamma$--X--M exhibit giant chiral splitting along M--$\Gamma$--M' in the absence of spin-orbit coupling. The chiral magnon modes have either left-handed or right-handed spin precession along a given N{\'e}el vector with opposite precessional angular momentum. Similar to the spin splitting in the electronic band structure, the chiral splitting of magnon bands in altermagnets even without relativistic effects differs altermagnets from ferromagnets and antiferromagnets\,\cite{Simejkal2023,Zhang2025a}, giving rise to novel applications in altermagnetic spintronics and magnonics. All magnon frequencies are real and non-negative with a global minimum at $\Gamma$, which, again, indicates a stable altermagnetic ground state\,\cite{Tellez-Mora2024}.

\begin{figure}
\centering
\includegraphics[width=\linewidth]{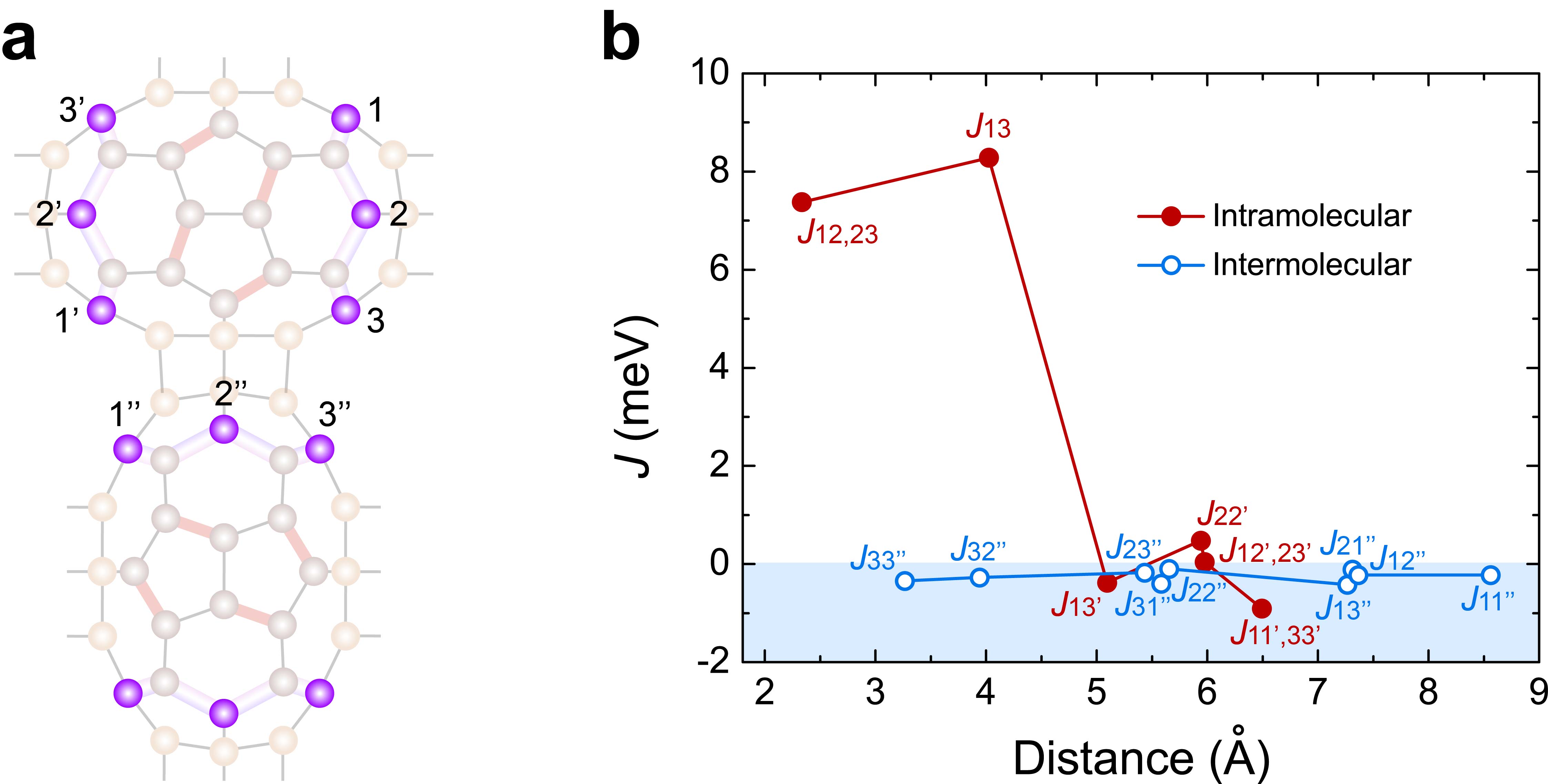}
\caption{
\textbf{Exchange interactions in altermagnetic C$_{40}$ monolayers.}
\textbf{a} Magnetic carbon atoms and \textbf{b} their corresponding exchange interactions as a function of distance.
}
\label{J} 
\end{figure}

\vspace{2mm}

\noindent\textbf{Exchange interactions.} 
We then examine the exchange interactions in altermagnetic C$_{40}$ monolayers. As shown in Fig.\,\ref{J}\textbf{a}, the three magnetic carbon atoms in the W-shaped chain share one $\pi$ electron, forming an effective spin-1/2 group. The exchange interactions between atoms $i$ and $j$ within one effective spin-1/2 group are denoted as $J_{ij}$, the intramolecular interactions between the two effective spin-1/2 groups are denoted as $J_{ij'}$, while the intermolecular interactions from the nearest neighbouring effective spin-1/2 groups are denoted as $J_{ij''}$. 

Figure\,\ref{J}\textbf{b} summarises the calculated $J$ parameters as a function of distance between these magnetic atoms. Within one effective spin-1/2 group, the interactions are ferromagnetic, with much stronger coupling strength ($J_{ij}>7$\,meV) than $J_{ij'}$ and $J_{ij''}$.

Intramolecular $J_{ij'}$ coupling is much smaller (between $-0.9$ and 0.5\,meV) than $J_{ij}$. The antiferromagnetic couplings ($J_{13'}=-0.38$\,meV, $J_{11'}=J_{33'}=-0.90$\,meV) are stronger than the ferromagnetic ones ($J_{22'}=0.48$\,meV, $J_{12'}=J_{23'}=0.04$\,meV). Interestingly, the long-range exchange interactions $J_{11'}$ and $J_{33'}$ from the largest intramolecular distance of 6.49\,\AA\ have the strongest coupling strength, whilst the smallest $J_{12'}$ and $J_{23'}$ have a relatively shorter distance of 5.98\,\AA.

The intermolecular interactions are all antiferromagnetic, with much weaker interaction strength than that of the intramolecular coupling. The largest intermolecular coupling comes from atom 1 and atom 3'' ($J_{13''}=-0.42$\,meV) at a distance of 7.27\,\AA, whereas the smallest intermolecular coupling corresponds to $J_{22''}=-0.10$\,meV at a smaller distance of 5.65\,\AA. The intermolecular interactions beyond the nearest neighbouring group quickly decay to zero.

\begin{figure*}
\centering
\includegraphics[width=\linewidth]{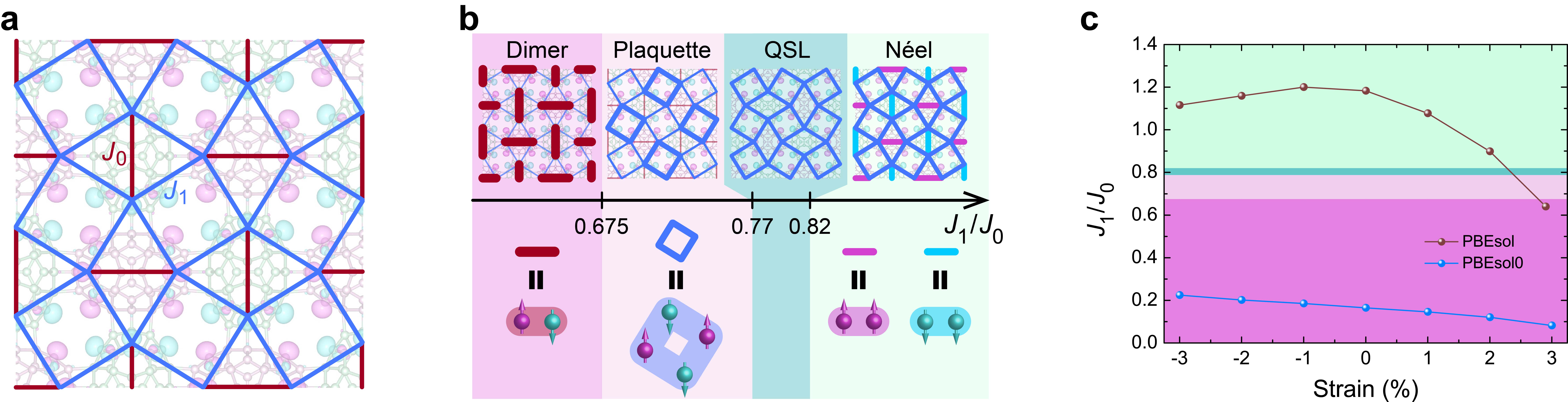}
\caption{
\textbf{Phase diagram of Shastry-Sutherland C$_{40}$ fullerene networks.}
\textbf{a} Shastry-Sutherland lattice, \textbf{b} phase diagram, and \textbf{c} order parameter $J_1/J_0$ as a function of bi-axial strain computed from semi-local PBEsol and hybrid-functional PBEsol0. 
}
\label{phase} 
\end{figure*}

\vspace{2mm}

\noindent\textbf{Phase diagram of the Shastry-Sutherland lattice.}
Having understood the exchange interactions of monolayer C$_{40}$ networks, we can build a lattice spin model by treating the three magnetic atoms in the effective spin-1/2 group as one spin-1/2 site, as these three atoms have dominant ferromagnetic interactions.

The effective spin-1/2 sites form the well-known Shastry-Sutherland lattice, as shown in Fig.\,\ref{phase}\textbf{a}. The effective exchange interactions between these effective spin-1/2 sites can be divided into two groups -- intramolecular exchange $J_0$ (marked in dark red) and intermolecular exchange $J_1$ (marked in blue). The effective exchange interactions can be defined as 
\begin{equation}
    J_\mathrm{eff}=\frac{\sum S^{(1)}_i J_{ij}^{(1,2)} S^{(2)}_j}{\sum S^{(1)}_i\sum S^{(2)}_j},
\end{equation}
where $S^{(1)}_i$ and $S^{(2)}_j$ indicate electron spin (taken as 1/6) from atomic site $i$ from effective spin-1/2 group 1 and atomic site $j$ from effective group 2 respectively, and $J_{ij}^{(1,2)}$ represents their exchange interactions. The intramolecular $J_0$ can be computed by choosing two effective spin-1/2 groups within the same molecule (i.e., $J_{ij'}$ in Fig.\,\ref{J}\textbf{a}), while the $J_1$ can be calculated from the intermolecular, nearest neighbouring effective spin-1/2 groups (i.e., $J_{ij''}$ in Fig.\,\ref{J}\textbf{a}). 

For monolayer C$_{40}$ networks, we obtain effective exchange interactions $J_0 = -0.21$\,meV and $J_1 = -0.25$\,meV, leading to a ratio $J_1/J_0$ of 1.18. According to the phase diagram for the Shastry-Sutherland lattice in Fig.\,\ref{phase}\,\textbf{b}, this corresponds to the N{\'e}el phase where $J_1$ dominates over $J_0$, and the Shastry-Sutherland lattice effectively reduces to the altermagnetic ground state as we find in \textit{ab initio} calculations. In the N{\'e}el state, the two spins within one C$_{40}$ unit are the same, whereas neighbouring fullerene molecules have opposite spins, which is exactly the ground-state altermagnetic phase.

The phase diagram can be tuned by the ratio between intermolecular and intramolecular coupling $J_1/J_0$. As shown in Fig.\,\ref{phase}\,\textbf{b}, decreasing the $J_1/J_0$ ratio to 0.82 
leads to a quantum spin liquid phase due to the frustrations between the intramolecular and intermolecular exchange couplings. This exotic phase has promising applications such as error-resistant topological qubits\,\cite{Klocke2024}. When the $J_1/J_0$ ratio lies between 0.77 and 0.675, a phase of independent ``plaquette'' units is formed, where each unit consists of four spins. This is known as the plaquette phase. In 2D C$_{40}$ networks, this phase corresponds to four spins located on four molecules around an interstice. Further decreasing the $J_1/J_0$ ratio below 0.675 results in the dimer valence bond solid phase. In the dimer phase, the two spins on each fullerene pair up into a singlet state as the intramolecular $J_0$ dominates, and the neighbouring singlet pairs stop to interact with each other.


To tune the ratios between $J_1$ and $J_0$, we introduce bi-axial strains.
Figure\,\ref{phase}\,\textbf{c} shows the $J_1/J_0$ ratios computed for monolayer C$_{40}$ networks under varied strain. 
For bi-axial strains between $-3$ and 2\%, the systems remain in the N{\'e}el order, indicating an altermagnetic ground state. When further increasing the strains to 3$\%$, the C$_{40}$ system passes through both the quantum spin liquid and plaquette phases, until reaching the dimer region. We use a strain step of 0.1\% between $2-3$\% strains. In the quantum spin liquid region, $J_1$ and $J_0$ become ill-defined, and our first-principles calculations show strong frustrations in the computed exchange interactions. In the plaquette phase, the four calculated $J_1$ terms around one spin-1/2 site are no longer symmetry equivalent but exhibit strong anisotropy, which agrees with the phase behaviours. Upon reaching the dimer phase, the four $J_1$ terms around one spin-1/2 site are equal again, leading to a well defined $J_1/J_0$ ratio.

Further unscreened hybrid functional calculations suggest that the dimer phase is stable at all strains between $\pm3\%$ in the zero-screening limit, which provides the lower bound for $J_1/J_0$ in free-standing monolayers without any substrates. This suggests that the phase transitions can be induced through control of screening effects such as the application of substrates with varied dielectric constants. 
Additionally, the dimer phase itself also exhibits rich physics such as triplon-bound state and spin-nematic ordering\,\cite{Zayed2017,McClarty2017,Wang2018b,Wulferding2021}, which can be further tuned by chemical decoration and exohedral doping. This will unlock future opportunities in realising room-temperature, organic quantum altermagnets to explore the complex many-body effects.



\vspace{2mm}

\noindent\textbf{Methods.} Density functional theory (DFT) calculations\,\cite{Hohenberg1964,Kohn1965} are performed using the Vienna \textit{ab initio} Simulation Package (\textsc{VASP})\,\cite{Kresse1996,Kresse1996a} under the generalised gradient approximation (GGA) formalism with projector augmented wave (PAW) basis sets\,\cite{Bloechl1994,Kresse1999}. The Perdew-Burke-Ernzerhof functional corrected for solids (PBEsol)\,\cite{Perdew2008} is used along with C $2s^22p^2$ as valence electrons.
A plane-wave cutoff of $800\,\mathrm{eV}$ is employed and the self-consistent field energy convergence criterion is set to be $10^{-6}\,\mathrm{eV}$. The Brillouin zone is sampled with a converged $\mathbf{k}$-mesh of $3\times3$. Spin polarisation is included throughout all the calculations. The crystal structures are fully relaxed using the conjugate gradient method\,\cite{Payne1992} until the Hellmann-Feynman forces are less than $10^{-2}\,\mathrm{eV/\AA}$. The vacuum spacing between monolayers is set to be more than 20\,\AA, and dipole corrections are employed perpendicular to the monolayers to eliminate interactions between periodic images\,\cite{Makov1995}. A Wannier tight-binding model\,\cite{Marzari1997,Souza2001,Marzari2012} is constructed using {\sc Wannier90}\,\cite{Mostofi2008,Mostofi2014,Pizzi2020} by projecting Wannier functions onto each $\sigma$ bond centre, with $sp^2$ carbon atoms having an extra Wannier function to mimic the $p_z$ orbitals. This bond-centre approach has been widely used in fullerene molecules\,\cite{Mostofi2008} and leads to 188 Wannier functions in a unit cell of two C$_{40}$ molecules. The exchange interactions are computed from the Green's function method\,\cite{Liechtenstein1987,Korotin2015} using the {\sc TB2J} package\,\cite{He2021} where the positions of Wannier centres and the magnetic moments are carefully checked. The magnon dispersion is computed under the Holstein-Primakoff transformation\,\cite{Holstein1940} based on the bosonic Hamiltonian\,\cite{Colpa1978}, as implemented in the \textsc{Magnopy} package that has been widely used to study low-dimensional antiferromagnets\,\cite{Rybakov2024,Boix-Constant2025}.
The chirality of magnons with band index $j$ is defined as the right-handed~(RH) or left-handed~(LH) spin precession along a given N{\'e}el vector with opposite precessional angular momentum\,\cite{Simejkal2023,Yuan2025}.
To analyse the effect of strain, the lattice parameters are modified (with respect to the fully relaxed structure) and fixed, while the atomic coordinates are allowed to fully relax. Unscreened hybrid functional PBEsol0 calculations\,\cite{Adamo1999} are also performed to study the exchange interactions in the zero-screening limit.

\noindent\textbf{Acknowledgment.} We thank Samzi Tishler at the University of Cambridge for proof reading. J.W. acknowledges support from the Cambridge Undergraduate Research Opportunities Programme and from Peterhouse for the James Porter Scholarship. 
B.P. acknowledges support from Magdalene College Cambridge for a Nevile Research Fellowship. The calculations were performed using resources provided by the Cambridge Service for Data Driven Discovery (CSD3) operated by the University of Cambridge Research Computing Service (\url{www.csd3.cam.ac.uk}), provided by Dell EMC and Intel using Tier-2 funding from the Engineering and Physical Sciences Research Council (capital grant EP/T022159/1), and DiRAC funding from the Science and Technology Facilities Council (\url{http://www.dirac.ac.uk}), as well as with computational support from the UK Materials and Molecular Modelling Hub, which is partially funded by EPSRC (EP/T022213/1, EP/W032260/1 and EP/P020194/1), for which access was obtained via the UKCP consortium and funded by EPSRC grant ref EP/P022561/1.

\bibliography{references}

\begin{thebibliography}{92}%
\makeatletter
\providecommand \@ifxundefined [1]{%
 \@ifx{#1\undefined}
}%
\providecommand \@ifnum [1]{%
 \ifnum #1\expandafter \@firstoftwo
 \else \expandafter \@secondoftwo
 \fi
}%
\providecommand \@ifx [1]{%
 \ifx #1\expandafter \@firstoftwo
 \else \expandafter \@secondoftwo
 \fi
}%
\providecommand \natexlab [1]{#1}%
\providecommand \enquote  [1]{``#1''}%
\providecommand \bibnamefont  [1]{#1}%
\providecommand \bibfnamefont [1]{#1}%
\providecommand \citenamefont [1]{#1}%
\providecommand \href@noop [0]{\@secondoftwo}%
\providecommand \href [0]{\begingroup \@sanitize@url \@href}%
\providecommand \@href[1]{\@@startlink{#1}\@@href}%
\providecommand \@@href[1]{\endgroup#1\@@endlink}%
\providecommand \@sanitize@url [0]{\catcode `\\12\catcode `\$12\catcode
  `\&12\catcode `\#12\catcode `\^12\catcode `\_12\catcode `\%12\relax}%
\providecommand \@@startlink[1]{}%
\providecommand \@@endlink[0]{}%
\providecommand \url  [0]{\begingroup\@sanitize@url \@url }%
\providecommand \@url [1]{\endgroup\@href {#1}{\urlprefix }}%
\providecommand \urlprefix  [0]{URL }%
\providecommand \Eprint [0]{\href }%
\providecommand \doibase [0]{http://dx.doi.org/}%
\providecommand \selectlanguage [0]{\@gobble}%
\providecommand \bibinfo  [0]{\@secondoftwo}%
\providecommand \bibfield  [0]{\@secondoftwo}%
\providecommand \translation [1]{[#1]}%
\providecommand \BibitemOpen [0]{}%
\providecommand \bibitemStop [0]{}%
\providecommand \bibitemNoStop [0]{.\EOS\space}%
\providecommand \EOS [0]{\spacefactor3000\relax}%
\providecommand \BibitemShut  [1]{\csname bibitem#1\endcsname}%
\let\auto@bib@innerbib\@empty
\bibitem [{\citenamefont {Sun}\ \emph {et~al.}(2025)\citenamefont {Sun},
  \citenamefont {Cao}, \citenamefont {Silveira}, \citenamefont {Fumega},
  \citenamefont {Hanindita}, \citenamefont {Ito}, \citenamefont {Lado},
  \citenamefont {Liljeroth}, \citenamefont {Foster},\ and\ \citenamefont
  {Kawai}}]{Sun2025}%
  \BibitemOpen
  \bibfield  {author} {\bibinfo {author} {\bibfnamefont {Kewei}\ \bibnamefont
  {Sun}}, \bibinfo {author} {\bibfnamefont {Nan}\ \bibnamefont {Cao}}, \bibinfo
  {author} {\bibfnamefont {Orlando~J.}\ \bibnamefont {Silveira}}, \bibinfo
  {author} {\bibfnamefont {Adolfo~O.}\ \bibnamefont {Fumega}}, \bibinfo
  {author} {\bibfnamefont {Fiona}\ \bibnamefont {Hanindita}}, \bibinfo {author}
  {\bibfnamefont {Shingo}\ \bibnamefont {Ito}}, \bibinfo {author}
  {\bibfnamefont {Jose~L.}\ \bibnamefont {Lado}}, \bibinfo {author}
  {\bibfnamefont {Peter}\ \bibnamefont {Liljeroth}}, \bibinfo {author}
  {\bibfnamefont {Adam~S.}\ \bibnamefont {Foster}}, \ and\ \bibinfo {author}
  {\bibfnamefont {Shigeki}\ \bibnamefont {Kawai}},\ }\bibfield  {title}
  {\enquote {\bibinfo {title} {On-surface synthesis of heisenberg spin-1/2
  antiferromagnetic molecular chains},}\ }\href {\doibase
  10.1126/sciadv.ads1641} {\bibfield  {journal} {\bibinfo  {journal} {Science
  Advances}\ }\textbf {\bibinfo {volume} {11}},\ \bibinfo {pages} {eads1641}
  (\bibinfo {year} {2025})}\BibitemShut {NoStop}%
\bibitem [{\citenamefont {Su}\ \emph {et~al.}(2025)\citenamefont {Su},
  \citenamefont {Ding}, \citenamefont {Hong}, \citenamefont {Ke}, \citenamefont
  {Yan}, \citenamefont {Li}, \citenamefont {Jiang},\ and\ \citenamefont
  {Yu}}]{Su2025}%
  \BibitemOpen
  \bibfield  {author} {\bibinfo {author} {\bibfnamefont {Xuelei}\ \bibnamefont
  {Su}}, \bibinfo {author} {\bibfnamefont {Zhihao}\ \bibnamefont {Ding}},
  \bibinfo {author} {\bibfnamefont {Ye}~\bibnamefont {Hong}}, \bibinfo {author}
  {\bibfnamefont {Nan}\ \bibnamefont {Ke}}, \bibinfo {author} {\bibfnamefont
  {KaKing}\ \bibnamefont {Yan}}, \bibinfo {author} {\bibfnamefont {Can}\
  \bibnamefont {Li}}, \bibinfo {author} {\bibfnamefont {Yi-Fan}\ \bibnamefont
  {Jiang}}, \ and\ \bibinfo {author} {\bibfnamefont {Ping}\ \bibnamefont
  {Yu}},\ }\bibfield  {title} {\enquote {\bibinfo {title} {Fabrication of
  spin-1/2 heisenberg antiferromagnetic chains via combined on-surface
  synthesis and reduction for spinon detection},}\ }\href {\doibase
  10.1038/s44160-025-00744-4} {\bibfield  {journal} {\bibinfo  {journal}
  {Nature Synthesis}\ }\textbf {\bibinfo {volume} {4}},\ \bibinfo {pages}
  {694--701} (\bibinfo {year} {2025})}\BibitemShut {NoStop}%
\bibitem [{\citenamefont {Fu}\ \emph {et~al.}(2025)\citenamefont {Fu},
  \citenamefont {Huang}, \citenamefont {Liu}, \citenamefont {Henriques},
  \citenamefont {Gao}, \citenamefont {Han}, \citenamefont {Chen}, \citenamefont
  {Wang}, \citenamefont {Palma}, \citenamefont {Cheng}, \citenamefont {Lin},
  \citenamefont {Du}, \citenamefont {Ma}, \citenamefont
  {Fern{\'a}ndez-Rossier}, \citenamefont {Feng},\ and\ \citenamefont
  {Gao}}]{Fu2025}%
  \BibitemOpen
  \bibfield  {author} {\bibinfo {author} {\bibfnamefont {Xiaoshuai}\
  \bibnamefont {Fu}}, \bibinfo {author} {\bibfnamefont {Li}~\bibnamefont
  {Huang}}, \bibinfo {author} {\bibfnamefont {Kun}\ \bibnamefont {Liu}},
  \bibinfo {author} {\bibfnamefont {Jo{\~a}o C.~G.}\ \bibnamefont {Henriques}},
  \bibinfo {author} {\bibfnamefont {Yixuan}\ \bibnamefont {Gao}}, \bibinfo
  {author} {\bibfnamefont {Xianghe}\ \bibnamefont {Han}}, \bibinfo {author}
  {\bibfnamefont {Hui}\ \bibnamefont {Chen}}, \bibinfo {author} {\bibfnamefont
  {Yan}\ \bibnamefont {Wang}}, \bibinfo {author} {\bibfnamefont
  {Carlos-Andres}\ \bibnamefont {Palma}}, \bibinfo {author} {\bibfnamefont
  {Zhihai}\ \bibnamefont {Cheng}}, \bibinfo {author} {\bibfnamefont {Xiao}\
  \bibnamefont {Lin}}, \bibinfo {author} {\bibfnamefont {Shixuan}\ \bibnamefont
  {Du}}, \bibinfo {author} {\bibfnamefont {Ji}~\bibnamefont {Ma}}, \bibinfo
  {author} {\bibfnamefont {Joaqu{\'i}n}\ \bibnamefont {Fern{\'a}ndez-Rossier}},
  \bibinfo {author} {\bibfnamefont {Xinliang}\ \bibnamefont {Feng}}, \ and\
  \bibinfo {author} {\bibfnamefont {Hong-Jun}\ \bibnamefont {Gao}},\ }\bibfield
   {title} {\enquote {\bibinfo {title} {Building spin-1/2 antiferromagnetic
  heisenberg chains with diaza-nanographenes},}\ }\href {\doibase
  10.1038/s44160-025-00743-5} {\bibfield  {journal} {\bibinfo  {journal}
  {Nature Synthesis}\ }\textbf {\bibinfo {volume} {4}},\ \bibinfo {pages}
  {684--693} (\bibinfo {year} {2025})}\BibitemShut {NoStop}%
\bibitem [{\citenamefont {Peng}\ and\ \citenamefont {Lu}(2025)}]{Peng2025b}%
  \BibitemOpen
  \bibfield  {author} {\bibinfo {author} {\bibfnamefont {Xinnan}\ \bibnamefont
  {Peng}}\ and\ \bibinfo {author} {\bibfnamefont {Jiong}\ \bibnamefont {Lu}},\
  }\bibfield  {title} {\enquote {\bibinfo {title} {Spin-1/2 heisenberg chains
  realized in $\pi$-electron systems},}\ }\href {\doibase
  10.1038/s44160-025-00757-z} {\bibfield  {journal} {\bibinfo  {journal}
  {Nature Synthesis}\ }\textbf {\bibinfo {volume} {4}},\ \bibinfo {pages}
  {668--670} (\bibinfo {year} {2025})}\BibitemShut {NoStop}%
\bibitem [{\citenamefont {Song}\ \emph {et~al.}(2025)\citenamefont {Song},
  \citenamefont {Teng}, \citenamefont {Tang}, \citenamefont {Xu}, \citenamefont
  {He}, \citenamefont {Ruan}, \citenamefont {Kojima}, \citenamefont {Hu},
  \citenamefont {Giessibl}, \citenamefont {Sakaguchi}, \citenamefont {Louie},\
  and\ \citenamefont {Lu}}]{Song2025}%
  \BibitemOpen
  \bibfield  {author} {\bibinfo {author} {\bibfnamefont {Shaotang}\
  \bibnamefont {Song}}, \bibinfo {author} {\bibfnamefont {Yu}~\bibnamefont
  {Teng}}, \bibinfo {author} {\bibfnamefont {Weichen}\ \bibnamefont {Tang}},
  \bibinfo {author} {\bibfnamefont {Zhen}\ \bibnamefont {Xu}}, \bibinfo
  {author} {\bibfnamefont {Yuanyuan}\ \bibnamefont {He}}, \bibinfo {author}
  {\bibfnamefont {Jiawei}\ \bibnamefont {Ruan}}, \bibinfo {author}
  {\bibfnamefont {Takahiro}\ \bibnamefont {Kojima}}, \bibinfo {author}
  {\bibfnamefont {Wenping}\ \bibnamefont {Hu}}, \bibinfo {author}
  {\bibfnamefont {Franz~J.}\ \bibnamefont {Giessibl}}, \bibinfo {author}
  {\bibfnamefont {Hiroshi}\ \bibnamefont {Sakaguchi}}, \bibinfo {author}
  {\bibfnamefont {Steven~G.}\ \bibnamefont {Louie}}, \ and\ \bibinfo {author}
  {\bibfnamefont {Jiong}\ \bibnamefont {Lu}},\ }\bibfield  {title} {\enquote
  {\bibinfo {title} {Janus graphene nanoribbons with localized states on a
  single zigzag edge},}\ }\href {\doibase 10.1038/s41586-024-08296-x}
  {\bibfield  {journal} {\bibinfo  {journal} {Nature}\ }\textbf {\bibinfo
  {volume} {637}},\ \bibinfo {pages} {580--586} (\bibinfo {year}
  {2025})}\BibitemShut {NoStop}%
\bibitem [{\citenamefont {Hayami}\ \emph {et~al.}(2019)\citenamefont {Hayami},
  \citenamefont {Yanagi},\ and\ \citenamefont {Kusunose}}]{Hayami2019}%
  \BibitemOpen
  \bibfield  {author} {\bibinfo {author} {\bibfnamefont {Satoru}\ \bibnamefont
  {Hayami}}, \bibinfo {author} {\bibfnamefont {Yuki}\ \bibnamefont {Yanagi}}, \
  and\ \bibinfo {author} {\bibfnamefont {Hiroaki}\ \bibnamefont {Kusunose}},\
  }\bibfield  {title} {\enquote {\bibinfo {title} {Momentum-dependent spin
  splitting by collinear antiferromagnetic ordering},}\ }\href {\doibase
  10.7566/JPSJ.88.123702} {\bibfield  {journal} {\bibinfo  {journal} {J. Phys.
  Soc. Jpn.}\ }\textbf {\bibinfo {volume} {88}},\ \bibinfo {pages} {123702}
  (\bibinfo {year} {2019})}\BibitemShut {NoStop}%
\bibitem [{\citenamefont {Yuan}\ \emph {et~al.}(2020)\citenamefont {Yuan},
  \citenamefont {Wang}, \citenamefont {Luo}, \citenamefont {Rashba},\ and\
  \citenamefont {Zunger}}]{Yuan2020}%
  \BibitemOpen
  \bibfield  {author} {\bibinfo {author} {\bibfnamefont {Lin-Ding}\
  \bibnamefont {Yuan}}, \bibinfo {author} {\bibfnamefont {Zhi}\ \bibnamefont
  {Wang}}, \bibinfo {author} {\bibfnamefont {Jun-Wei}\ \bibnamefont {Luo}},
  \bibinfo {author} {\bibfnamefont {Emmanuel~I.}\ \bibnamefont {Rashba}}, \
  and\ \bibinfo {author} {\bibfnamefont {Alex}\ \bibnamefont {Zunger}},\
  }\bibfield  {title} {\enquote {\bibinfo {title} {Giant momentum-dependent
  spin splitting in centrosymmetric low-$z$ antiferromagnets},}\ }\href
  {\doibase 10.1103/PhysRevB.102.014422} {\bibfield  {journal} {\bibinfo
  {journal} {Phys. Rev. B}\ }\textbf {\bibinfo {volume} {102}},\ \bibinfo
  {pages} {014422} (\bibinfo {year} {2020})}\BibitemShut {NoStop}%
\bibitem [{\citenamefont {{\v{S}}mejkal}\ \emph {et~al.}(2020)\citenamefont
  {{\v{S}}mejkal}, \citenamefont {Gonz{\'a}lez-Hern{\'a}ndez}, \citenamefont
  {Jungwirth},\ and\ \citenamefont {Sinova}}]{Smejkal2020}%
  \BibitemOpen
  \bibfield  {author} {\bibinfo {author} {\bibfnamefont {Libor}\ \bibnamefont
  {{\v{S}}mejkal}}, \bibinfo {author} {\bibfnamefont {Rafael}\ \bibnamefont
  {Gonz{\'a}lez-Hern{\'a}ndez}}, \bibinfo {author} {\bibfnamefont
  {T.}~\bibnamefont {Jungwirth}}, \ and\ \bibinfo {author} {\bibfnamefont
  {J.}~\bibnamefont {Sinova}},\ }\bibfield  {title} {\enquote {\bibinfo {title}
  {Crystal time-reversal symmetry breaking and spontaneous hall effect in
  collinear antiferromagnets},}\ }\href {\doibase 10.1126/sciadv.aaz8809}
  {\bibfield  {journal} {\bibinfo  {journal} {Science Advances}\ }\textbf
  {\bibinfo {volume} {6}},\ \bibinfo {pages} {eaaz8809} (\bibinfo {year}
  {2020})}\BibitemShut {NoStop}%
\bibitem [{\citenamefont {Ma}\ \emph {et~al.}(2021)\citenamefont {Ma},
  \citenamefont {Hu}, \citenamefont {Li}, \citenamefont {Liu}, \citenamefont
  {Yao}, \citenamefont {Jia},\ and\ \citenamefont {Liu}}]{Ma2021a}%
  \BibitemOpen
  \bibfield  {author} {\bibinfo {author} {\bibfnamefont {Hai-Yang}\
  \bibnamefont {Ma}}, \bibinfo {author} {\bibfnamefont {Mengli}\ \bibnamefont
  {Hu}}, \bibinfo {author} {\bibfnamefont {Nana}\ \bibnamefont {Li}}, \bibinfo
  {author} {\bibfnamefont {Jianpeng}\ \bibnamefont {Liu}}, \bibinfo {author}
  {\bibfnamefont {Wang}\ \bibnamefont {Yao}}, \bibinfo {author} {\bibfnamefont
  {Jin-Feng}\ \bibnamefont {Jia}}, \ and\ \bibinfo {author} {\bibfnamefont
  {Junwei}\ \bibnamefont {Liu}},\ }\bibfield  {title} {\enquote {\bibinfo
  {title} {Multifunctional antiferromagnetic materials with giant
  piezomagnetism and noncollinear spin current},}\ }\href {\doibase
  10.1038/s41467-021-23127-7} {\bibfield  {journal} {\bibinfo  {journal}
  {Nature Communications}\ }\textbf {\bibinfo {volume} {12}},\ \bibinfo {pages}
  {2846} (\bibinfo {year} {2021})}\BibitemShut {NoStop}%
\bibitem [{\citenamefont {Mazin}\ \emph {et~al.}(2021)\citenamefont {Mazin},
  \citenamefont {Koepernik}, \citenamefont {Johannes}, \citenamefont
  {Gonz{\'a}lez-Hern{\'a}ndez},\ and\ \citenamefont
  {{\v{S}}mejkal}}]{Mazin2021}%
  \BibitemOpen
  \bibfield  {author} {\bibinfo {author} {\bibfnamefont {Igor~I.}\ \bibnamefont
  {Mazin}}, \bibinfo {author} {\bibfnamefont {Klaus}\ \bibnamefont
  {Koepernik}}, \bibinfo {author} {\bibfnamefont {Michelle~D.}\ \bibnamefont
  {Johannes}}, \bibinfo {author} {\bibfnamefont {Rafael}\ \bibnamefont
  {Gonz{\'a}lez-Hern{\'a}ndez}}, \ and\ \bibinfo {author} {\bibfnamefont
  {Libor}\ \bibnamefont {{\v{S}}mejkal}},\ }\bibfield  {title} {\enquote
  {\bibinfo {title} {{{Prediction of unconventional magnetism in doped
  FeSb$_2$}}},}\ }\href {\doibase 10.1073/pnas.2108924118} {\bibfield
  {journal} {\bibinfo  {journal} {Proc Natl Acad Sci}\ }\textbf {\bibinfo
  {volume} {118}},\ \bibinfo {pages} {e2108924118} (\bibinfo {year}
  {2021})}\BibitemShut {NoStop}%
\bibitem [{\citenamefont {{\v{S}}mejkal}\ \emph
  {et~al.}(2022{\natexlab{a}})\citenamefont {{\v{S}}mejkal}, \citenamefont
  {Sinova},\ and\ \citenamefont {Jungwirth}}]{Simejkal2022}%
  \BibitemOpen
  \bibfield  {author} {\bibinfo {author} {\bibfnamefont {Libor}\ \bibnamefont
  {{\v{S}}mejkal}}, \bibinfo {author} {\bibfnamefont {Jairo}\ \bibnamefont
  {Sinova}}, \ and\ \bibinfo {author} {\bibfnamefont {Tomas}\ \bibnamefont
  {Jungwirth}},\ }\bibfield  {title} {\enquote {\bibinfo {title} {Emerging
  research landscape of altermagnetism},}\ }\href {\doibase
  10.1103/PhysRevX.12.040501} {\bibfield  {journal} {\bibinfo  {journal} {Phys.
  Rev. X}\ }\textbf {\bibinfo {volume} {12}},\ \bibinfo {pages} {040501}
  (\bibinfo {year} {2022}{\natexlab{a}})}\BibitemShut {NoStop}%
\bibitem [{\citenamefont {{\v{S}}mejkal}\ \emph
  {et~al.}(2022{\natexlab{b}})\citenamefont {{\v{S}}mejkal}, \citenamefont
  {Sinova},\ and\ \citenamefont {Jungwirth}}]{Simejkal2022a}%
  \BibitemOpen
  \bibfield  {author} {\bibinfo {author} {\bibfnamefont {Libor}\ \bibnamefont
  {{\v{S}}mejkal}}, \bibinfo {author} {\bibfnamefont {Jairo}\ \bibnamefont
  {Sinova}}, \ and\ \bibinfo {author} {\bibfnamefont {Tomas}\ \bibnamefont
  {Jungwirth}},\ }\bibfield  {title} {\enquote {\bibinfo {title} {Beyond
  conventional ferromagnetism and antiferromagnetism: A phase with
  nonrelativistic spin and crystal rotation symmetry},}\ }\href {\doibase
  10.1103/PhysRevX.12.031042} {\bibfield  {journal} {\bibinfo  {journal} {Phys.
  Rev. X}\ }\textbf {\bibinfo {volume} {12}},\ \bibinfo {pages} {031042}
  (\bibinfo {year} {2022}{\natexlab{b}})}\BibitemShut {NoStop}%
\bibitem [{\citenamefont {Fedchenko}\ \emph {et~al.}(2024)\citenamefont
  {Fedchenko}, \citenamefont {Min{\'a}r}, \citenamefont {Akashdeep},
  \citenamefont {D'souza}, \citenamefont {Vasilyev}, \citenamefont {Tkach},
  \citenamefont {Odenbreit}, \citenamefont {Nguyen}, \citenamefont
  {Kutnyakhov}, \citenamefont {Wind}, \citenamefont {Wenthaus}, \citenamefont
  {Scholz}, \citenamefont {Rossnagel}, \citenamefont {Hoesch}, \citenamefont
  {Aeschlimann}, \citenamefont {Stadtm{\"u}ller}, \citenamefont {Kl{\"a}ui},
  \citenamefont {Sch{\"o}nhense}, \citenamefont {Jungwirth}, \citenamefont
  {Hellenes}, \citenamefont {Jakob}, \citenamefont {{\v{S}}mejkal},
  \citenamefont {Sinova},\ and\ \citenamefont {Elmers}}]{Fedchenko2024}%
  \BibitemOpen
  \bibfield  {author} {\bibinfo {author} {\bibfnamefont {Olena}\ \bibnamefont
  {Fedchenko}}, \bibinfo {author} {\bibfnamefont {Jan}\ \bibnamefont
  {Min{\'a}r}}, \bibinfo {author} {\bibfnamefont {Akashdeep}\ \bibnamefont
  {Akashdeep}}, \bibinfo {author} {\bibfnamefont {Sunil~Wilfred}\ \bibnamefont
  {D'souza}}, \bibinfo {author} {\bibfnamefont {Dmitry}\ \bibnamefont
  {Vasilyev}}, \bibinfo {author} {\bibfnamefont {Olena}\ \bibnamefont {Tkach}},
  \bibinfo {author} {\bibfnamefont {Lukas}\ \bibnamefont {Odenbreit}}, \bibinfo
  {author} {\bibfnamefont {Quynh}\ \bibnamefont {Nguyen}}, \bibinfo {author}
  {\bibfnamefont {Dmytro}\ \bibnamefont {Kutnyakhov}}, \bibinfo {author}
  {\bibfnamefont {Nils}\ \bibnamefont {Wind}}, \bibinfo {author} {\bibfnamefont
  {Lukas}\ \bibnamefont {Wenthaus}}, \bibinfo {author} {\bibfnamefont {Markus}\
  \bibnamefont {Scholz}}, \bibinfo {author} {\bibfnamefont {Kai}\ \bibnamefont
  {Rossnagel}}, \bibinfo {author} {\bibfnamefont {Moritz}\ \bibnamefont
  {Hoesch}}, \bibinfo {author} {\bibfnamefont {Martin}\ \bibnamefont
  {Aeschlimann}}, \bibinfo {author} {\bibfnamefont {Benjamin}\ \bibnamefont
  {Stadtm{\"u}ller}}, \bibinfo {author} {\bibfnamefont {Mathias}\ \bibnamefont
  {Kl{\"a}ui}}, \bibinfo {author} {\bibfnamefont {Gerd}\ \bibnamefont
  {Sch{\"o}nhense}}, \bibinfo {author} {\bibfnamefont {Tomas}\ \bibnamefont
  {Jungwirth}}, \bibinfo {author} {\bibfnamefont {Anna~Birk}\ \bibnamefont
  {Hellenes}}, \bibinfo {author} {\bibfnamefont {Gerhard}\ \bibnamefont
  {Jakob}}, \bibinfo {author} {\bibfnamefont {Libor}\ \bibnamefont
  {{\v{S}}mejkal}}, \bibinfo {author} {\bibfnamefont {Jairo}\ \bibnamefont
  {Sinova}}, \ and\ \bibinfo {author} {\bibfnamefont {Hans-Joachim}\
  \bibnamefont {Elmers}},\ }\bibfield  {title} {\enquote {\bibinfo {title}
  {{{Observation of time-reversal symmetry breaking in the band structure of
  altermagnetic RuO$_2$}}},}\ }\href {\doibase 10.1126/sciadv.adj4883}
  {\bibfield  {journal} {\bibinfo  {journal} {Science Advances}\ }\textbf
  {\bibinfo {volume} {10}},\ \bibinfo {pages} {eadj4883} (\bibinfo {year}
  {2024})}\BibitemShut {NoStop}%
\bibitem [{\citenamefont {Krempask{\'y}}\ \emph {et~al.}(2024)\citenamefont
  {Krempask{\'y}}, \citenamefont {{\v{S}}mejkal}, \citenamefont {D'Souza},
  \citenamefont {Hajlaoui}, \citenamefont {Springholz}, \citenamefont
  {Uhl{\'i}{\vr}ov{\'a}}, \citenamefont {Alarab}, \citenamefont {Constantinou},
  \citenamefont {Strocov}, \citenamefont {Usanov}, \citenamefont {Pudelko},
  \citenamefont {Gonz{\'a}lez-Hern{\'a}ndez}, \citenamefont {Birk~Hellenes},
  \citenamefont {Jansa}, \citenamefont {Reichlov{\'a}}, \citenamefont
  {{\v{S}}ob{\'a}{\v{n}}}, \citenamefont {Gonzalez~Betancourt}, \citenamefont
  {Wadley}, \citenamefont {Sinova}, \citenamefont {Kriegner}, \citenamefont
  {Min{\'a}r}, \citenamefont {Dil},\ and\ \citenamefont
  {Jungwirth}}]{Krempasky2024}%
  \BibitemOpen
  \bibfield  {author} {\bibinfo {author} {\bibfnamefont {J.}~\bibnamefont
  {Krempask{\'y}}}, \bibinfo {author} {\bibfnamefont {L.}~\bibnamefont
  {{\v{S}}mejkal}}, \bibinfo {author} {\bibfnamefont {S.~W.}\ \bibnamefont
  {D'Souza}}, \bibinfo {author} {\bibfnamefont {M.}~\bibnamefont {Hajlaoui}},
  \bibinfo {author} {\bibfnamefont {G.}~\bibnamefont {Springholz}}, \bibinfo
  {author} {\bibfnamefont {K.}~\bibnamefont {Uhl{\'i}{\vr}ov{\'a}}}, \bibinfo
  {author} {\bibfnamefont {F.}~\bibnamefont {Alarab}}, \bibinfo {author}
  {\bibfnamefont {P.~C.}\ \bibnamefont {Constantinou}}, \bibinfo {author}
  {\bibfnamefont {V.}~\bibnamefont {Strocov}}, \bibinfo {author} {\bibfnamefont
  {D.}~\bibnamefont {Usanov}}, \bibinfo {author} {\bibfnamefont {W.~R.}\
  \bibnamefont {Pudelko}}, \bibinfo {author} {\bibfnamefont {R.}~\bibnamefont
  {Gonz{\'a}lez-Hern{\'a}ndez}}, \bibinfo {author} {\bibfnamefont
  {A.}~\bibnamefont {Birk~Hellenes}}, \bibinfo {author} {\bibfnamefont
  {Z.}~\bibnamefont {Jansa}}, \bibinfo {author} {\bibfnamefont
  {H.}~\bibnamefont {Reichlov{\'a}}}, \bibinfo {author} {\bibfnamefont
  {Z.}~\bibnamefont {{\v{S}}ob{\'a}{\v{n}}}}, \bibinfo {author} {\bibfnamefont
  {R.~D.}\ \bibnamefont {Gonzalez~Betancourt}}, \bibinfo {author}
  {\bibfnamefont {P.}~\bibnamefont {Wadley}}, \bibinfo {author} {\bibfnamefont
  {J.}~\bibnamefont {Sinova}}, \bibinfo {author} {\bibfnamefont
  {D.}~\bibnamefont {Kriegner}}, \bibinfo {author} {\bibfnamefont
  {J.}~\bibnamefont {Min{\'a}r}}, \bibinfo {author} {\bibfnamefont {J.~H.}\
  \bibnamefont {Dil}}, \ and\ \bibinfo {author} {\bibfnamefont
  {T.}~\bibnamefont {Jungwirth}},\ }\bibfield  {title} {\enquote {\bibinfo
  {title} {Altermagnetic lifting of kramers spin degeneracy},}\ }\href
  {\doibase 10.1038/s41586-023-06907-7} {\bibfield  {journal} {\bibinfo
  {journal} {Nature}\ }\textbf {\bibinfo {volume} {626}},\ \bibinfo {pages}
  {517--522} (\bibinfo {year} {2024})}\BibitemShut {NoStop}%
\bibitem [{\citenamefont {Bose}\ \emph {et~al.}(2024)\citenamefont {Bose},
  \citenamefont {Vadnais},\ and\ \citenamefont {Paramekanti}}]{Bose2024}%
  \BibitemOpen
  \bibfield  {author} {\bibinfo {author} {\bibfnamefont {Anjishnu}\
  \bibnamefont {Bose}}, \bibinfo {author} {\bibfnamefont {Samuel}\ \bibnamefont
  {Vadnais}}, \ and\ \bibinfo {author} {\bibfnamefont {Arun}\ \bibnamefont
  {Paramekanti}},\ }\bibfield  {title} {\enquote {\bibinfo {title}
  {Altermagnetism and superconductivity in a multiorbital $t\ensuremath{-}j$
  model},}\ }\href {\doibase 10.1103/PhysRevB.110.205120} {\bibfield  {journal}
  {\bibinfo  {journal} {Phys. Rev. B}\ }\textbf {\bibinfo {volume} {110}},\
  \bibinfo {pages} {205120} (\bibinfo {year} {2024})}\BibitemShut {NoStop}%
\bibitem [{\citenamefont {Pupim}\ and\ \citenamefont
  {Scheurer}(2025)}]{Pupim2025}%
  \BibitemOpen
  \bibfield  {author} {\bibinfo {author} {\bibfnamefont {Lucas~V.}\
  \bibnamefont {Pupim}}\ and\ \bibinfo {author} {\bibfnamefont {Mathias~S.}\
  \bibnamefont {Scheurer}},\ }\bibfield  {title} {\enquote {\bibinfo {title}
  {Adatom engineering magnetic order in superconductors: Applications to
  altermagnetic superconductivity},}\ }\href {\doibase
  10.1103/PhysRevLett.134.146001} {\bibfield  {journal} {\bibinfo  {journal}
  {Phys. Rev. Lett.}\ }\textbf {\bibinfo {volume} {134}},\ \bibinfo {pages}
  {146001} (\bibinfo {year} {2025})}\BibitemShut {NoStop}%
\bibitem [{\citenamefont {Sobral}\ \emph {et~al.}(2025)\citenamefont {Sobral},
  \citenamefont {Mandal},\ and\ \citenamefont {Scheurer}}]{Sobral2025}%
  \BibitemOpen
  \bibfield  {author} {\bibinfo {author} {\bibfnamefont {Jo\~ao~Augusto}\
  \bibnamefont {Sobral}}, \bibinfo {author} {\bibfnamefont {Subrata}\
  \bibnamefont {Mandal}}, \ and\ \bibinfo {author} {\bibfnamefont {Mathias~S.}\
  \bibnamefont {Scheurer}},\ }\bibfield  {title} {\enquote {\bibinfo {title}
  {Fractionalized altermagnets: From neighboring and altermagnetic spin liquids
  to spin-symmetric band splitting},}\ }\href {\doibase
  10.1103/PhysRevResearch.7.023152} {\bibfield  {journal} {\bibinfo  {journal}
  {Phys. Rev. Res.}\ }\textbf {\bibinfo {volume} {7}},\ \bibinfo {pages}
  {023152} (\bibinfo {year} {2025})}\BibitemShut {NoStop}%
\bibitem [{\citenamefont {Sato}\ \emph {et~al.}(2024)\citenamefont {Sato},
  \citenamefont {Haddad}, \citenamefont {Fulga}, \citenamefont {Assaad},\ and\
  \citenamefont {van~den Brink}}]{Sato2024}%
  \BibitemOpen
  \bibfield  {author} {\bibinfo {author} {\bibfnamefont {Toshihiro}\
  \bibnamefont {Sato}}, \bibinfo {author} {\bibfnamefont {Sonia}\ \bibnamefont
  {Haddad}}, \bibinfo {author} {\bibfnamefont {Ion~Cosma}\ \bibnamefont
  {Fulga}}, \bibinfo {author} {\bibfnamefont {Fakher~F.}\ \bibnamefont
  {Assaad}}, \ and\ \bibinfo {author} {\bibfnamefont {Jeroen}\ \bibnamefont
  {van~den Brink}},\ }\bibfield  {title} {\enquote {\bibinfo {title}
  {Altermagnetic anomalous hall effect emerging from electronic
  correlations},}\ }\href {\doibase 10.1103/PhysRevLett.133.086503} {\bibfield
  {journal} {\bibinfo  {journal} {Phys. Rev. Lett.}\ }\textbf {\bibinfo
  {volume} {133}},\ \bibinfo {pages} {086503} (\bibinfo {year}
  {2024})}\BibitemShut {NoStop}%
\bibitem [{\citenamefont {Das}\ \emph {et~al.}(2024)\citenamefont {Das},
  \citenamefont {Leeb}, \citenamefont {Knolle},\ and\ \citenamefont
  {Knap}}]{Das2024}%
  \BibitemOpen
  \bibfield  {author} {\bibinfo {author} {\bibfnamefont {Purnendu}\
  \bibnamefont {Das}}, \bibinfo {author} {\bibfnamefont {Valentin}\
  \bibnamefont {Leeb}}, \bibinfo {author} {\bibfnamefont {Johannes}\
  \bibnamefont {Knolle}}, \ and\ \bibinfo {author} {\bibfnamefont {Michael}\
  \bibnamefont {Knap}},\ }\bibfield  {title} {\enquote {\bibinfo {title}
  {Realizing altermagnetism in fermi-hubbard models with ultracold atoms},}\
  }\href {\doibase 10.1103/PhysRevLett.132.263402} {\bibfield  {journal}
  {\bibinfo  {journal} {Phys. Rev. Lett.}\ }\textbf {\bibinfo {volume} {132}},\
  \bibinfo {pages} {263402} (\bibinfo {year} {2024})}\BibitemShut {NoStop}%
\bibitem [{\citenamefont {Giuli}\ \emph {et~al.}(2025)\citenamefont {Giuli},
  \citenamefont {Mejuto-Zaera},\ and\ \citenamefont {Capone}}]{Giuli2025}%
  \BibitemOpen
  \bibfield  {author} {\bibinfo {author} {\bibfnamefont {Samuele}\ \bibnamefont
  {Giuli}}, \bibinfo {author} {\bibfnamefont {Carlos}\ \bibnamefont
  {Mejuto-Zaera}}, \ and\ \bibinfo {author} {\bibfnamefont {Massimo}\
  \bibnamefont {Capone}},\ }\bibfield  {title} {\enquote {\bibinfo {title}
  {Altermagnetism from interaction-driven itinerant magnetism},}\ }\href
  {\doibase 10.1103/PhysRevB.111.L020401} {\bibfield  {journal} {\bibinfo
  {journal} {Phys. Rev. B}\ }\textbf {\bibinfo {volume} {111}},\ \bibinfo
  {pages} {L020401} (\bibinfo {year} {2025})}\BibitemShut {NoStop}%
\bibitem [{\citenamefont {He}\ \emph {et~al.}(2025)\citenamefont {He},
  \citenamefont {Zhao}, \citenamefont {Luo},\ and\ \citenamefont
  {Hu}}]{He2025}%
  \BibitemOpen
  \bibfield  {author} {\bibinfo {author} {\bibfnamefont {Saisai}\ \bibnamefont
  {He}}, \bibinfo {author} {\bibfnamefont {Jize}\ \bibnamefont {Zhao}},
  \bibinfo {author} {\bibfnamefont {Hong-Gang}\ \bibnamefont {Luo}}, \ and\
  \bibinfo {author} {\bibfnamefont {Shijie}\ \bibnamefont {Hu}},\ }\bibfield
  {title} {\enquote {\bibinfo {title} {Altermagnetism and beyond in the
  $t\text{\ensuremath{-}}{t}^{\ensuremath{'}}\text{\ensuremath{-}}\ensuremath{\delta}$
  fermi-hubbard model},}\ }\href {\doibase 10.1103/4mv8-tb66} {\bibfield
  {journal} {\bibinfo  {journal} {Phys. Rev. B}\ }\textbf {\bibinfo {volume}
  {112}},\ \bibinfo {pages} {035108} (\bibinfo {year} {2025})}\BibitemShut
  {NoStop}%
\bibitem [{\citenamefont {Zhao}\ \emph {et~al.}(2025)\citenamefont {Zhao},
  \citenamefont {Yang}, \citenamefont {Guo}, \citenamefont {Luo},\ and\
  \citenamefont {Zhong}}]{Zhao2025}%
  \BibitemOpen
  \bibfield  {author} {\bibinfo {author} {\bibfnamefont {Miaomiao}\
  \bibnamefont {Zhao}}, \bibinfo {author} {\bibfnamefont {Wei-Wei}\
  \bibnamefont {Yang}}, \bibinfo {author} {\bibfnamefont {Xueming}\
  \bibnamefont {Guo}}, \bibinfo {author} {\bibfnamefont {Hong-Gang}\
  \bibnamefont {Luo}}, \ and\ \bibinfo {author} {\bibfnamefont {Yin}\
  \bibnamefont {Zhong}},\ }\bibfield  {title} {\enquote {\bibinfo {title}
  {Altermagnetism in heavy-fermion systems: Mean-field study on the kondo
  lattice},}\ }\href {\doibase 10.1103/PhysRevB.111.085145} {\bibfield
  {journal} {\bibinfo  {journal} {Phys. Rev. B}\ }\textbf {\bibinfo {volume}
  {111}},\ \bibinfo {pages} {085145} (\bibinfo {year} {2025})}\BibitemShut
  {NoStop}%
\bibitem [{\citenamefont {Sriram~Shastry}\ and\ \citenamefont
  {Sutherland}(1981)}]{SriramShastry1981}%
  \BibitemOpen
  \bibfield  {author} {\bibinfo {author} {\bibfnamefont {B.}~\bibnamefont
  {Sriram~Shastry}}\ and\ \bibinfo {author} {\bibfnamefont {Bill}\ \bibnamefont
  {Sutherland}},\ }\bibfield  {title} {\enquote {\bibinfo {title} {Exact ground
  state of a quantum mechanical antiferromagnet},}\ }\href {\doibase
  10.1016/0378-4363(81)90838-X} {\bibfield  {journal} {\bibinfo  {journal}
  {Physica B+C}\ }\textbf {\bibinfo {volume} {108}},\ \bibinfo {pages}
  {1069--1070} (\bibinfo {year} {1981})}\BibitemShut {NoStop}%
\bibitem [{\citenamefont {Miyahara}\ and\ \citenamefont
  {Ueda}(1999)}]{Miyahara1999}%
  \BibitemOpen
  \bibfield  {author} {\bibinfo {author} {\bibfnamefont {Shin}\ \bibnamefont
  {Miyahara}}\ and\ \bibinfo {author} {\bibfnamefont {Kazuo}\ \bibnamefont
  {Ueda}},\ }\bibfield  {title} {\enquote {\bibinfo {title} {Exact dimer ground
  state of the two dimensional heisenberg spin system
  ${\mathrm{srcu}}_{2}({\mathrm{bo}}_{3}){}_{2}$},}\ }\href {\doibase
  10.1103/PhysRevLett.82.3701} {\bibfield  {journal} {\bibinfo  {journal}
  {Phys. Rev. Lett.}\ }\textbf {\bibinfo {volume} {82}},\ \bibinfo {pages}
  {3701--3704} (\bibinfo {year} {1999})}\BibitemShut {NoStop}%
\bibitem [{\citenamefont {Knetter}\ \emph {et~al.}(2000)\citenamefont
  {Knetter}, \citenamefont {B\"uhler}, \citenamefont {M\"uller-Hartmann},\ and\
  \citenamefont {Uhrig}}]{Knetter2000}%
  \BibitemOpen
  \bibfield  {author} {\bibinfo {author} {\bibfnamefont {Christian}\
  \bibnamefont {Knetter}}, \bibinfo {author} {\bibfnamefont {Alexander}\
  \bibnamefont {B\"uhler}}, \bibinfo {author} {\bibfnamefont {Erwin}\
  \bibnamefont {M\"uller-Hartmann}}, \ and\ \bibinfo {author} {\bibfnamefont
  {G\"otz~S.}\ \bibnamefont {Uhrig}},\ }\bibfield  {title} {\enquote {\bibinfo
  {title} {Dispersion and symmetry of bound states in the shastry-sutherland
  model},}\ }\href {\doibase 10.1103/PhysRevLett.85.3958} {\bibfield  {journal}
  {\bibinfo  {journal} {Phys. Rev. Lett.}\ }\textbf {\bibinfo {volume} {85}},\
  \bibinfo {pages} {3958--3961} (\bibinfo {year} {2000})}\BibitemShut {NoStop}%
\bibitem [{\citenamefont {Koga}\ and\ \citenamefont
  {Kawakami}(2000)}]{Koga2000}%
  \BibitemOpen
  \bibfield  {author} {\bibinfo {author} {\bibfnamefont {Akihisa}\ \bibnamefont
  {Koga}}\ and\ \bibinfo {author} {\bibfnamefont {Norio}\ \bibnamefont
  {Kawakami}},\ }\bibfield  {title} {\enquote {\bibinfo {title} {Quantum phase
  transitions in the shastry-sutherland model for
  ${\mathrm{srcu}}_{2}({\mathrm{bo}}_{3}{)}_{2}$},}\ }\href {\doibase
  10.1103/PhysRevLett.84.4461} {\bibfield  {journal} {\bibinfo  {journal}
  {Phys. Rev. Lett.}\ }\textbf {\bibinfo {volume} {84}},\ \bibinfo {pages}
  {4461--4464} (\bibinfo {year} {2000})}\BibitemShut {NoStop}%
\bibitem [{\citenamefont {L\"auchli}\ \emph {et~al.}(2002)\citenamefont
  {L\"auchli}, \citenamefont {Wessel},\ and\ \citenamefont
  {Sigrist}}]{Laeuchli2002}%
  \BibitemOpen
  \bibfield  {author} {\bibinfo {author} {\bibfnamefont {Andreas}\ \bibnamefont
  {L\"auchli}}, \bibinfo {author} {\bibfnamefont {Stefan}\ \bibnamefont
  {Wessel}}, \ and\ \bibinfo {author} {\bibfnamefont {Manfred}\ \bibnamefont
  {Sigrist}},\ }\bibfield  {title} {\enquote {\bibinfo {title} {Phase diagram
  of the quadrumerized shastry-sutherland model},}\ }\href {\doibase
  10.1103/PhysRevB.66.014401} {\bibfield  {journal} {\bibinfo  {journal} {Phys.
  Rev. B}\ }\textbf {\bibinfo {volume} {66}},\ \bibinfo {pages} {014401}
  (\bibinfo {year} {2002})}\BibitemShut {NoStop}%
\bibitem [{\citenamefont {Dorier}\ \emph {et~al.}(2008)\citenamefont {Dorier},
  \citenamefont {Schmidt},\ and\ \citenamefont {Mila}}]{Dorier2008}%
  \BibitemOpen
  \bibfield  {author} {\bibinfo {author} {\bibfnamefont {J.}~\bibnamefont
  {Dorier}}, \bibinfo {author} {\bibfnamefont {K.~P.}\ \bibnamefont {Schmidt}},
  \ and\ \bibinfo {author} {\bibfnamefont {F.}~\bibnamefont {Mila}},\
  }\bibfield  {title} {\enquote {\bibinfo {title} {Theory of magnetization
  plateaux in the shastry-sutherland model},}\ }\href {\doibase
  10.1103/PhysRevLett.101.250402} {\bibfield  {journal} {\bibinfo  {journal}
  {Phys. Rev. Lett.}\ }\textbf {\bibinfo {volume} {101}},\ \bibinfo {pages}
  {250402} (\bibinfo {year} {2008})}\BibitemShut {NoStop}%
\bibitem [{\citenamefont {Brassington}\ \emph {et~al.}(2024)\citenamefont
  {Brassington}, \citenamefont {Ma}, \citenamefont {Sala}, \citenamefont
  {Kolesnikov}, \citenamefont {Taddei}, \citenamefont {Wu}, \citenamefont
  {Choi}, \citenamefont {Wang}, \citenamefont {Xie}, \citenamefont {Ma},
  \citenamefont {Zhou},\ and\ \citenamefont {Aczel}}]{Brassington2024}%
  \BibitemOpen
  \bibfield  {author} {\bibinfo {author} {\bibfnamefont {A.}~\bibnamefont
  {Brassington}}, \bibinfo {author} {\bibfnamefont {Q.}~\bibnamefont {Ma}},
  \bibinfo {author} {\bibfnamefont {G.}~\bibnamefont {Sala}}, \bibinfo {author}
  {\bibfnamefont {A.~I.}\ \bibnamefont {Kolesnikov}}, \bibinfo {author}
  {\bibfnamefont {K.~M.}\ \bibnamefont {Taddei}}, \bibinfo {author}
  {\bibfnamefont {Y.}~\bibnamefont {Wu}}, \bibinfo {author} {\bibfnamefont
  {E.~S.}\ \bibnamefont {Choi}}, \bibinfo {author} {\bibfnamefont
  {H.}~\bibnamefont {Wang}}, \bibinfo {author} {\bibfnamefont {W.}~\bibnamefont
  {Xie}}, \bibinfo {author} {\bibfnamefont {J.}~\bibnamefont {Ma}}, \bibinfo
  {author} {\bibfnamefont {H.~D.}\ \bibnamefont {Zhou}}, \ and\ \bibinfo
  {author} {\bibfnamefont {A.~A.}\ \bibnamefont {Aczel}},\ }\bibfield  {title}
  {\enquote {\bibinfo {title} {Magnetic properties of the quasi-xy
  shastry-sutherland magnet
  ${\mathrm{er}}_{2}{\mathrm{be}}_{2}{\mathrm{sio}}_{7}$},}\ }\href {\doibase
  10.1103/PhysRevMaterials.8.094001} {\bibfield  {journal} {\bibinfo  {journal}
  {Phys. Rev. Mater.}\ }\textbf {\bibinfo {volume} {8}},\ \bibinfo {pages}
  {094001} (\bibinfo {year} {2024})}\BibitemShut {NoStop}%
\bibitem [{\citenamefont {Ferrari}\ and\ \citenamefont
  {Valent\'{\i}}(2024)}]{Ferrari2024}%
  \BibitemOpen
  \bibfield  {author} {\bibinfo {author} {\bibfnamefont {Francesco}\
  \bibnamefont {Ferrari}}\ and\ \bibinfo {author} {\bibfnamefont {Roser}\
  \bibnamefont {Valent\'{\i}}},\ }\bibfield  {title} {\enquote {\bibinfo
  {title} {Altermagnetism on the shastry-sutherland lattice},}\ }\href
  {\doibase 10.1103/PhysRevB.110.205140} {\bibfield  {journal} {\bibinfo
  {journal} {Phys. Rev. B}\ }\textbf {\bibinfo {volume} {110}},\ \bibinfo
  {pages} {205140} (\bibinfo {year} {2024})}\BibitemShut {NoStop}%
\bibitem [{\citenamefont {Chung}\ and\ \citenamefont {Kim}(2004)}]{Chung2004}%
  \BibitemOpen
  \bibfield  {author} {\bibinfo {author} {\bibfnamefont {Chung-Hou}\
  \bibnamefont {Chung}}\ and\ \bibinfo {author} {\bibfnamefont {Yong~Baek}\
  \bibnamefont {Kim}},\ }\bibfield  {title} {\enquote {\bibinfo {title}
  {Competing orders and superconductivity in the doped mott insulator on the
  shastry-sutherland lattice},}\ }\href {\doibase
  10.1103/PhysRevLett.93.207004} {\bibfield  {journal} {\bibinfo  {journal}
  {Phys. Rev. Lett.}\ }\textbf {\bibinfo {volume} {93}},\ \bibinfo {pages}
  {207004} (\bibinfo {year} {2004})}\BibitemShut {NoStop}%
\bibitem [{\citenamefont {Yang}\ \emph {et~al.}(2008)\citenamefont {Yang},
  \citenamefont {Kim}, \citenamefont {Yu},\ and\ \citenamefont
  {Park}}]{Yang2008b}%
  \BibitemOpen
  \bibfield  {author} {\bibinfo {author} {\bibfnamefont {Bohm-Jung}\
  \bibnamefont {Yang}}, \bibinfo {author} {\bibfnamefont {Yong~Baek}\
  \bibnamefont {Kim}}, \bibinfo {author} {\bibfnamefont {Jaejun}\ \bibnamefont
  {Yu}}, \ and\ \bibinfo {author} {\bibfnamefont {Kwon}\ \bibnamefont {Park}},\
  }\bibfield  {title} {\enquote {\bibinfo {title} {Doped valence-bond solid and
  superconductivity on the shastry-sutherland lattice},}\ }\href {\doibase
  10.1103/PhysRevB.77.104507} {\bibfield  {journal} {\bibinfo  {journal} {Phys.
  Rev. B}\ }\textbf {\bibinfo {volume} {77}},\ \bibinfo {pages} {104507}
  (\bibinfo {year} {2008})}\BibitemShut {NoStop}%
\bibitem [{\citenamefont {Liu}\ \emph {et~al.}(2007)\citenamefont {Liu},
  \citenamefont {Trivedi}, \citenamefont {Lee}, \citenamefont {Harmon},\ and\
  \citenamefont {Schmalian}}]{Liu2007}%
  \BibitemOpen
  \bibfield  {author} {\bibinfo {author} {\bibfnamefont {Jun}\ \bibnamefont
  {Liu}}, \bibinfo {author} {\bibfnamefont {Nandini}\ \bibnamefont {Trivedi}},
  \bibinfo {author} {\bibfnamefont {Yongbin}\ \bibnamefont {Lee}}, \bibinfo
  {author} {\bibfnamefont {B.~N.}\ \bibnamefont {Harmon}}, \ and\ \bibinfo
  {author} {\bibfnamefont {J\"org}\ \bibnamefont {Schmalian}},\ }\bibfield
  {title} {\enquote {\bibinfo {title} {Quantum phases in a doped mott insulator
  on the shastry-sutherland lattice},}\ }\href {\doibase
  10.1103/PhysRevLett.99.227003} {\bibfield  {journal} {\bibinfo  {journal}
  {Phys. Rev. Lett.}\ }\textbf {\bibinfo {volume} {99}},\ \bibinfo {pages}
  {227003} (\bibinfo {year} {2007})}\BibitemShut {NoStop}%
\bibitem [{\citenamefont {Lee}\ \emph {et~al.}(2019)\citenamefont {Lee},
  \citenamefont {You}, \citenamefont {Sachdev},\ and\ \citenamefont
  {Vishwanath}}]{Lee2019}%
  \BibitemOpen
  \bibfield  {author} {\bibinfo {author} {\bibfnamefont {Jong~Yeon}\
  \bibnamefont {Lee}}, \bibinfo {author} {\bibfnamefont {Yi-Zhuang}\
  \bibnamefont {You}}, \bibinfo {author} {\bibfnamefont {Subir}\ \bibnamefont
  {Sachdev}}, \ and\ \bibinfo {author} {\bibfnamefont {Ashvin}\ \bibnamefont
  {Vishwanath}},\ }\bibfield  {title} {\enquote {\bibinfo {title} {Signatures
  of a deconfined phase transition on the shastry-sutherland lattice:
  Applications to quantum critical
  ${\mathrm{srcu}}_{2}({\mathrm{bo}}_{3}{)}_{2}$},}\ }\href {\doibase
  10.1103/PhysRevX.9.041037} {\bibfield  {journal} {\bibinfo  {journal} {Phys.
  Rev. X}\ }\textbf {\bibinfo {volume} {9}},\ \bibinfo {pages} {041037}
  (\bibinfo {year} {2019})}\BibitemShut {NoStop}%
\bibitem [{\citenamefont {Shi}\ \emph {et~al.}(2022)\citenamefont {Shi},
  \citenamefont {Dissanayake}, \citenamefont {Corboz}, \citenamefont
  {Steinhardt}, \citenamefont {Graf}, \citenamefont {Silevitch}, \citenamefont
  {Dabkowska}, \citenamefont {Rosenbaum}, \citenamefont {Mila},\ and\
  \citenamefont {Haravifard}}]{Shi2022}%
  \BibitemOpen
  \bibfield  {author} {\bibinfo {author} {\bibfnamefont {Zhenzhong}\
  \bibnamefont {Shi}}, \bibinfo {author} {\bibfnamefont {Sachith}\ \bibnamefont
  {Dissanayake}}, \bibinfo {author} {\bibfnamefont {Philippe}\ \bibnamefont
  {Corboz}}, \bibinfo {author} {\bibfnamefont {William}\ \bibnamefont
  {Steinhardt}}, \bibinfo {author} {\bibfnamefont {David}\ \bibnamefont
  {Graf}}, \bibinfo {author} {\bibfnamefont {D.~M.}\ \bibnamefont {Silevitch}},
  \bibinfo {author} {\bibfnamefont {Hanna~A.}\ \bibnamefont {Dabkowska}},
  \bibinfo {author} {\bibfnamefont {T.~F.}\ \bibnamefont {Rosenbaum}}, \bibinfo
  {author} {\bibfnamefont {Fr{\'e}d{\'e}ric}\ \bibnamefont {Mila}}, \ and\
  \bibinfo {author} {\bibfnamefont {Sara}\ \bibnamefont {Haravifard}},\
  }\bibfield  {title} {\enquote {\bibinfo {title} {Discovery of quantum phases
  in the shastry-sutherland compound srcu2(bo3)2 under extreme conditions of
  field and pressure},}\ }\href {\doibase 10.1038/s41467-022-30036-w}
  {\bibfield  {journal} {\bibinfo  {journal} {Nature Communications}\ }\textbf
  {\bibinfo {volume} {13}},\ \bibinfo {pages} {2301} (\bibinfo {year}
  {2022})}\BibitemShut {NoStop}%
\bibitem [{\citenamefont {Pula}\ \emph {et~al.}(2024)\citenamefont {Pula},
  \citenamefont {Sharma}, \citenamefont {Gautreau}, \citenamefont {K.~P.},
  \citenamefont {Kanigel}, \citenamefont {Frontzek}, \citenamefont {Dolling},
  \citenamefont {Clark}, \citenamefont {Dunsiger}, \citenamefont {Ghara},\ and\
  \citenamefont {Luke}}]{Pula2024}%
  \BibitemOpen
  \bibfield  {author} {\bibinfo {author} {\bibfnamefont {M.}~\bibnamefont
  {Pula}}, \bibinfo {author} {\bibfnamefont {S.}~\bibnamefont {Sharma}},
  \bibinfo {author} {\bibfnamefont {J.}~\bibnamefont {Gautreau}}, \bibinfo
  {author} {\bibfnamefont {Sajilesh}\ \bibnamefont {K.~P.}}, \bibinfo {author}
  {\bibfnamefont {A.}~\bibnamefont {Kanigel}}, \bibinfo {author} {\bibfnamefont
  {M.~D.}\ \bibnamefont {Frontzek}}, \bibinfo {author} {\bibfnamefont {T.~N.}\
  \bibnamefont {Dolling}}, \bibinfo {author} {\bibfnamefont {L.}~\bibnamefont
  {Clark}}, \bibinfo {author} {\bibfnamefont {S.}~\bibnamefont {Dunsiger}},
  \bibinfo {author} {\bibfnamefont {A.}~\bibnamefont {Ghara}}, \ and\ \bibinfo
  {author} {\bibfnamefont {G.~M.}\ \bibnamefont {Luke}},\ }\bibfield  {title}
  {\enquote {\bibinfo {title} {Candidate for a quantum spin liquid ground state
  in the shastry-sutherland lattice material
  ${\mathrm{yb}}_{2}{\mathrm{be}}_{2}{\mathrm{geo}}_{7}$},}\ }\href {\doibase
  10.1103/PhysRevB.110.014412} {\bibfield  {journal} {\bibinfo  {journal}
  {Phys. Rev. B}\ }\textbf {\bibinfo {volume} {110}},\ \bibinfo {pages}
  {014412} (\bibinfo {year} {2024})}\BibitemShut {NoStop}%
\bibitem [{\citenamefont {Corboz}\ \emph {et~al.}(2025)\citenamefont {Corboz},
  \citenamefont {Zhang}, \citenamefont {Ponsioen},\ and\ \citenamefont
  {Mila}}]{Corboz2025}%
  \BibitemOpen
  \bibfield  {author} {\bibinfo {author} {\bibfnamefont {Philippe}\
  \bibnamefont {Corboz}}, \bibinfo {author} {\bibfnamefont {Yining}\
  \bibnamefont {Zhang}}, \bibinfo {author} {\bibfnamefont {Boris}\ \bibnamefont
  {Ponsioen}}, \ and\ \bibinfo {author} {\bibfnamefont {Fr{\'e}d{\'e}ric}\
  \bibnamefont {Mila}},\ }\bibfield  {title} {\enquote {\bibinfo {title}
  {Quantum spin liquid phase in the shastry-sutherland model revealed by
  high-precision infinite projected entangled-pair states},}\ }\href@noop {}
  {\bibfield  {journal} {\bibinfo  {journal} {arXiv:2502.14091}\ } (\bibinfo
  {year} {2025})}\BibitemShut {NoStop}%
\bibitem [{\citenamefont {Yazyev}(2010)}]{Yazyev2010}%
  \BibitemOpen
  \bibfield  {author} {\bibinfo {author} {\bibfnamefont {Oleg~V}\ \bibnamefont
  {Yazyev}},\ }\bibfield  {title} {\enquote {\bibinfo {title} {Emergence of
  magnetism in graphene materials and nanostructures},}\ }\href {\doibase
  10.1088/0034-4885/73/5/056501} {\bibfield  {journal} {\bibinfo  {journal}
  {Rep. Prog. Phys.}\ }\textbf {\bibinfo {volume} {73}},\ \bibinfo {pages}
  {056501} (\bibinfo {year} {2010})}\BibitemShut {NoStop}%
\bibitem [{\citenamefont {Slota}\ \emph {et~al.}(2018)\citenamefont {Slota},
  \citenamefont {Keerthi}, \citenamefont {Myers}, \citenamefont {Tretyakov},
  \citenamefont {Baumgarten}, \citenamefont {Ardavan}, \citenamefont {Sadeghi},
  \citenamefont {Lambert}, \citenamefont {Narita}, \citenamefont {M{\"u}llen},\
  and\ \citenamefont {Bogani}}]{Slota2018}%
  \BibitemOpen
  \bibfield  {author} {\bibinfo {author} {\bibfnamefont {Michael}\ \bibnamefont
  {Slota}}, \bibinfo {author} {\bibfnamefont {Ashok}\ \bibnamefont {Keerthi}},
  \bibinfo {author} {\bibfnamefont {William~K.}\ \bibnamefont {Myers}},
  \bibinfo {author} {\bibfnamefont {Evgeny}\ \bibnamefont {Tretyakov}},
  \bibinfo {author} {\bibfnamefont {Martin}\ \bibnamefont {Baumgarten}},
  \bibinfo {author} {\bibfnamefont {Arzhang}\ \bibnamefont {Ardavan}}, \bibinfo
  {author} {\bibfnamefont {Hatef}\ \bibnamefont {Sadeghi}}, \bibinfo {author}
  {\bibfnamefont {Colin~J.}\ \bibnamefont {Lambert}}, \bibinfo {author}
  {\bibfnamefont {Akimitsu}\ \bibnamefont {Narita}}, \bibinfo {author}
  {\bibfnamefont {Klaus}\ \bibnamefont {M{\"u}llen}}, \ and\ \bibinfo {author}
  {\bibfnamefont {Lapo}\ \bibnamefont {Bogani}},\ }\bibfield  {title} {\enquote
  {\bibinfo {title} {Magnetic edge states and coherent manipulation of graphene
  nanoribbons},}\ }\href {\doibase 10.1038/s41586-018-0154-7} {\bibfield
  {journal} {\bibinfo  {journal} {Nature}\ }\textbf {\bibinfo {volume} {557}},\
  \bibinfo {pages} {691--695} (\bibinfo {year} {2018})}\BibitemShut {NoStop}%
\bibitem [{\citenamefont {Ma}\ \emph {et~al.}(2025)\citenamefont {Ma},
  \citenamefont {Tepliakov}, \citenamefont {Mostofi},\ and\ \citenamefont
  {Pizzochero}}]{Ma2025}%
  \BibitemOpen
  \bibfield  {author} {\bibinfo {author} {\bibfnamefont {Ruize}\ \bibnamefont
  {Ma}}, \bibinfo {author} {\bibfnamefont {Nikita~V.}\ \bibnamefont
  {Tepliakov}}, \bibinfo {author} {\bibfnamefont {Arash~A.}\ \bibnamefont
  {Mostofi}}, \ and\ \bibinfo {author} {\bibfnamefont {Michele}\ \bibnamefont
  {Pizzochero}},\ }\bibfield  {title} {\enquote {\bibinfo {title} {Electrically
  tunable ultraflat bands and $\pi$-electron magnetism in graphene
  nanoribbons},}\ }\href {\doibase 10.1021/acs.jpclett.5c00121} {\bibfield
  {journal} {\bibinfo  {journal} {J. Phys. Chem. Lett.}\ }\textbf {\bibinfo
  {volume} {16}},\ \bibinfo {pages} {1680--1685} (\bibinfo {year}
  {2025})}\BibitemShut {NoStop}%
\bibitem [{\citenamefont {Wu}\ \emph {et~al.}(2025)\citenamefont {Wu},
  \citenamefont {Pingen},\ and\ \citenamefont {Peng}}]{Wu2025a}%
  \BibitemOpen
  \bibfield  {author} {\bibinfo {author} {\bibfnamefont {Jiaqi}\ \bibnamefont
  {Wu}}, \bibinfo {author} {\bibfnamefont {Leonard~Werner}\ \bibnamefont
  {Pingen}}, \ and\ \bibinfo {author} {\bibfnamefont {Bo}~\bibnamefont
  {Peng}},\ }\bibfield  {title} {\enquote {\bibinfo {title} {Symmetry-induced
  magnetism in fullerene monolayers},}\ }\href@noop {} {\bibfield  {journal}
  {\bibinfo  {journal} {arXiv:2508.18125}\ } (\bibinfo {year}
  {2025})}\BibitemShut {NoStop}%
\bibitem [{\citenamefont {Peng}\ and\ \citenamefont
  {Pizzochero}(2025{\natexlab{a}})}]{Peng2025d}%
  \BibitemOpen
  \bibfield  {author} {\bibinfo {author} {\bibfnamefont {Bo}~\bibnamefont
  {Peng}}\ and\ \bibinfo {author} {\bibfnamefont {Michele}\ \bibnamefont
  {Pizzochero}},\ }\bibfield  {title} {\enquote {\bibinfo {title} {Designing
  antiferromagnetic spin-1/2 chains in janus fullerene nanoribbons},}\
  }\href@noop {} {\bibfield  {journal} {\bibinfo  {journal} {arXiv:2508.18849}\
  } (\bibinfo {year} {2025}{\natexlab{a}})}\BibitemShut {NoStop}%
\bibitem [{\citenamefont {Pingen}\ \emph {et~al.}(2025)\citenamefont {Pingen},
  \citenamefont {Wu},\ and\ \citenamefont {Peng}}]{Pingen2025}%
  \BibitemOpen
  \bibfield  {author} {\bibinfo {author} {\bibfnamefont {Leonard~Werner}\
  \bibnamefont {Pingen}}, \bibinfo {author} {\bibfnamefont {Jiaqi}\
  \bibnamefont {Wu}}, \ and\ \bibinfo {author} {\bibfnamefont {Bo}~\bibnamefont
  {Peng}},\ }\bibfield  {title} {\enquote {\bibinfo {title} {Tunable quantum
  anomalous hall effect in fullerene monolayers},}\ }\href@noop {} {\bibfield
  {journal} {\bibinfo  {journal} {arXiv:2508.19849}\ } (\bibinfo {year}
  {2025})}\BibitemShut {NoStop}%
\bibitem [{\citenamefont {Hou}\ \emph {et~al.}(2022)\citenamefont {Hou},
  \citenamefont {Cui}, \citenamefont {Guan}, \citenamefont {Wang},
  \citenamefont {Li}, \citenamefont {Liu}, \citenamefont {Zhu},\ and\
  \citenamefont {Zheng}}]{Hou2022}%
  \BibitemOpen
  \bibfield  {author} {\bibinfo {author} {\bibfnamefont {Lingxiang}\
  \bibnamefont {Hou}}, \bibinfo {author} {\bibfnamefont {Xueping}\ \bibnamefont
  {Cui}}, \bibinfo {author} {\bibfnamefont {Bo}~\bibnamefont {Guan}}, \bibinfo
  {author} {\bibfnamefont {Shaozhi}\ \bibnamefont {Wang}}, \bibinfo {author}
  {\bibfnamefont {Ruian}\ \bibnamefont {Li}}, \bibinfo {author} {\bibfnamefont
  {Yunqi}\ \bibnamefont {Liu}}, \bibinfo {author} {\bibfnamefont {Daoben}\
  \bibnamefont {Zhu}}, \ and\ \bibinfo {author} {\bibfnamefont {Jian}\
  \bibnamefont {Zheng}},\ }\bibfield  {title} {\enquote {\bibinfo {title}
  {Synthesis of a monolayer fullerene network},}\ }\href {\doibase
  10.1038/s41586-022-04771-5} {\bibfield  {journal} {\bibinfo  {journal}
  {Nature}\ }\textbf {\bibinfo {volume} {606}},\ \bibinfo {pages} {507--510}
  (\bibinfo {year} {2022})}\BibitemShut {NoStop}%
\bibitem [{\citenamefont {Peng}(2022)}]{Peng2022c}%
  \BibitemOpen
  \bibfield  {author} {\bibinfo {author} {\bibfnamefont {Bo}~\bibnamefont
  {Peng}},\ }\bibfield  {title} {\enquote {\bibinfo {title} {Monolayer
  fullerene networks as photocatalysts for overall water splitting},}\ }\href
  {\doibase 10.1021/jacs.2c08054} {\bibfield  {journal} {\bibinfo  {journal}
  {J. Am. Chem. Soc.}\ }\textbf {\bibinfo {volume} {144}},\ \bibinfo {pages}
  {19921--19931} (\bibinfo {year} {2022})}\BibitemShut {NoStop}%
\bibitem [{\citenamefont {Peng}(2023)}]{Peng2023}%
  \BibitemOpen
  \bibfield  {author} {\bibinfo {author} {\bibfnamefont {Bo}~\bibnamefont
  {Peng}},\ }\bibfield  {title} {\enquote {\bibinfo {title} {{{Stability and
  Strength of Monolayer Polymeric C$_{60}$}}},}\ }\href {\doibase
  10.1021/acs.nanolett.2c04497} {\bibfield  {journal} {\bibinfo  {journal}
  {Nano Lett.}\ }\textbf {\bibinfo {volume} {23}},\ \bibinfo {pages} {652--658}
  (\bibinfo {year} {2023})}\BibitemShut {NoStop}%
\bibitem [{\citenamefont {Jones}\ and\ \citenamefont {Peng}(2023)}]{Jones2023}%
  \BibitemOpen
  \bibfield  {author} {\bibinfo {author} {\bibfnamefont {Cory}\ \bibnamefont
  {Jones}}\ and\ \bibinfo {author} {\bibfnamefont {Bo}~\bibnamefont {Peng}},\
  }\bibfield  {title} {\enquote {\bibinfo {title} {Boosting photocatalytic
  water splitting of polymeric c60 by reduced dimensionality from
  two-dimensional monolayer to one-dimensional chain},}\ }\href {\doibase
  10.1021/acs.jpclett.3c02578} {\bibfield  {journal} {\bibinfo  {journal} {J.
  Phys. Chem. Lett.}\ }\textbf {\bibinfo {volume} {14}},\ \bibinfo {pages}
  {11768--11773} (\bibinfo {year} {2023})}\BibitemShut {NoStop}%
\bibitem [{\citenamefont {Wu}\ and\ \citenamefont {Peng}(2025)}]{Wu2025}%
  \BibitemOpen
  \bibfield  {author} {\bibinfo {author} {\bibfnamefont {Jiaqi}\ \bibnamefont
  {Wu}}\ and\ \bibinfo {author} {\bibfnamefont {Bo}~\bibnamefont {Peng}},\
  }\bibfield  {title} {\enquote {\bibinfo {title} {Smallest [5,6]fullerene as
  building blocks for 2d networks with superior stability and enhanced
  photocatalytic performance},}\ }\href {\doibase 10.1021/jacs.4c13167}
  {\bibfield  {journal} {\bibinfo  {journal} {J. Am. Chem. Soc.}\ }\textbf
  {\bibinfo {volume} {147}},\ \bibinfo {pages} {1749--1757} (\bibinfo {year}
  {2025})}\BibitemShut {NoStop}%
\bibitem [{\citenamefont {Shearsby}\ \emph {et~al.}(2025)\citenamefont
  {Shearsby}, \citenamefont {Wu}, \citenamefont {Yang},\ and\ \citenamefont
  {Peng}}]{Shearsby2025}%
  \BibitemOpen
  \bibfield  {author} {\bibinfo {author} {\bibfnamefont {Dylan}\ \bibnamefont
  {Shearsby}}, \bibinfo {author} {\bibfnamefont {Jiaqi}\ \bibnamefont {Wu}},
  \bibinfo {author} {\bibfnamefont {Dekun}\ \bibnamefont {Yang}}, \ and\
  \bibinfo {author} {\bibfnamefont {Bo}~\bibnamefont {Peng}},\ }\bibfield
  {title} {\enquote {\bibinfo {title} {Tuning electronic and optical properties
  of 2d polymeric c60 by stacking two layers},}\ }\href {\doibase
  10.1039/D4NR04540H} {\bibfield  {journal} {\bibinfo  {journal} {Nanoscale}\
  }\textbf {\bibinfo {volume} {17}},\ \bibinfo {pages} {2616--2620} (\bibinfo
  {year} {2025})}\BibitemShut {NoStop}%
\bibitem [{\citenamefont {Kayley}\ and\ \citenamefont
  {Peng}(2025)}]{Kayley2025}%
  \BibitemOpen
  \bibfield  {author} {\bibinfo {author} {\bibfnamefont {Darius}\ \bibnamefont
  {Kayley}}\ and\ \bibinfo {author} {\bibfnamefont {Bo}~\bibnamefont {Peng}},\
  }\bibfield  {title} {\enquote {\bibinfo {title} {C$_{60}$ building blocks
  with tuneable structures for tailored functionalities},}\ }\href {\doibase
  10.1016/j.commt.2025.100030} {\bibfield  {journal} {\bibinfo  {journal}
  {Computational Materials Today}\ }\textbf {\bibinfo {volume} {6}},\ \bibinfo
  {pages} {100030} (\bibinfo {year} {2025})}\BibitemShut {NoStop}%
\bibitem [{\citenamefont {Shaikh}\ and\ \citenamefont
  {Peng}(2025)}]{Shaikh2025}%
  \BibitemOpen
  \bibfield  {author} {\bibinfo {author} {\bibfnamefont {Armaan}\ \bibnamefont
  {Shaikh}}\ and\ \bibinfo {author} {\bibfnamefont {Bo}~\bibnamefont {Peng}},\
  }\bibfield  {title} {\enquote {\bibinfo {title} {Negative and positive
  anisotropic thermal expansion in 2d fullerene networks},}\ }\href@noop {}
  {\bibfield  {journal} {\bibinfo  {journal} {arXiv:}\ ,\ \bibinfo {pages}
  {2504.02037}} (\bibinfo {year} {2025})}\BibitemShut {NoStop}%
\bibitem [{\citenamefont {Peng}\ and\ \citenamefont
  {Pizzochero}(2025{\natexlab{b}})}]{Peng2025c}%
  \BibitemOpen
  \bibfield  {author} {\bibinfo {author} {\bibfnamefont {Bo}~\bibnamefont
  {Peng}}\ and\ \bibinfo {author} {\bibfnamefont {Michele}\ \bibnamefont
  {Pizzochero}},\ }\bibfield  {title} {\enquote {\bibinfo {title} {Electronic
  structure of fullerene nanoribbons},}\ }\href {\doibase
  10.1021/acsnano.5c08991} {\bibfield  {journal} {\bibinfo  {journal} {ACS
  Nano}\ }\textbf {\bibinfo {volume} {19}},\ \bibinfo {pages} {29637--29645}
  (\bibinfo {year} {2025}{\natexlab{b}})}\BibitemShut {NoStop}%
\bibitem [{\citenamefont {Peng}\ and\ \citenamefont
  {Pizzochero}(2025{\natexlab{c}})}]{Peng2025a}%
  \BibitemOpen
  \bibfield  {author} {\bibinfo {author} {\bibfnamefont {Bo}~\bibnamefont
  {Peng}}\ and\ \bibinfo {author} {\bibfnamefont {Michele}\ \bibnamefont
  {Pizzochero}},\ }\bibfield  {title} {\enquote {\bibinfo {title} {{{Monolayer
  C$_{60}$ networks: a first-principles perspective}}},}\ }\href {\doibase
  10.1039/D5CC02473K} {\bibfield  {journal} {\bibinfo  {journal} {Chem.
  Commun.}\ }\textbf {\bibinfo {volume} {61}},\ \bibinfo {pages} {10287--10302}
  (\bibinfo {year} {2025}{\natexlab{c}})}\BibitemShut {NoStop}%
\bibitem [{\citenamefont {Meirzadeh}\ \emph {et~al.}(2023)\citenamefont
  {Meirzadeh}, \citenamefont {Evans}, \citenamefont {Rezaee}, \citenamefont
  {Milich}, \citenamefont {Dionne}, \citenamefont {Darlington}, \citenamefont
  {Bao}, \citenamefont {Bartholomew}, \citenamefont {Handa}, \citenamefont
  {Rizzo}, \citenamefont {Wiscons}, \citenamefont {Reza}, \citenamefont
  {Zangiabadi}, \citenamefont {Fardian-Melamed}, \citenamefont {Crowther},
  \citenamefont {Schuck}, \citenamefont {Basov}, \citenamefont {Zhu},
  \citenamefont {Giri}, \citenamefont {Hopkins}, \citenamefont {Kim},
  \citenamefont {Steigerwald}, \citenamefont {Yang}, \citenamefont {Nuckolls},\
  and\ \citenamefont {Roy}}]{Meirzadeh2023}%
  \BibitemOpen
  \bibfield  {author} {\bibinfo {author} {\bibfnamefont {Elena}\ \bibnamefont
  {Meirzadeh}}, \bibinfo {author} {\bibfnamefont {Austin~M.}\ \bibnamefont
  {Evans}}, \bibinfo {author} {\bibfnamefont {Mehdi}\ \bibnamefont {Rezaee}},
  \bibinfo {author} {\bibfnamefont {Milena}\ \bibnamefont {Milich}}, \bibinfo
  {author} {\bibfnamefont {Connor~J.}\ \bibnamefont {Dionne}}, \bibinfo
  {author} {\bibfnamefont {Thomas~P.}\ \bibnamefont {Darlington}}, \bibinfo
  {author} {\bibfnamefont {Si~Tong}\ \bibnamefont {Bao}}, \bibinfo {author}
  {\bibfnamefont {Amymarie~K.}\ \bibnamefont {Bartholomew}}, \bibinfo {author}
  {\bibfnamefont {Taketo}\ \bibnamefont {Handa}}, \bibinfo {author}
  {\bibfnamefont {Daniel~J.}\ \bibnamefont {Rizzo}}, \bibinfo {author}
  {\bibfnamefont {Ren~A.}\ \bibnamefont {Wiscons}}, \bibinfo {author}
  {\bibfnamefont {Mahniz}\ \bibnamefont {Reza}}, \bibinfo {author}
  {\bibfnamefont {Amirali}\ \bibnamefont {Zangiabadi}}, \bibinfo {author}
  {\bibfnamefont {Natalie}\ \bibnamefont {Fardian-Melamed}}, \bibinfo {author}
  {\bibfnamefont {Andrew~C.}\ \bibnamefont {Crowther}}, \bibinfo {author}
  {\bibfnamefont {P.~James}\ \bibnamefont {Schuck}}, \bibinfo {author}
  {\bibfnamefont {D.~N.}\ \bibnamefont {Basov}}, \bibinfo {author}
  {\bibfnamefont {Xiaoyang}\ \bibnamefont {Zhu}}, \bibinfo {author}
  {\bibfnamefont {Ashutosh}\ \bibnamefont {Giri}}, \bibinfo {author}
  {\bibfnamefont {Patrick~E.}\ \bibnamefont {Hopkins}}, \bibinfo {author}
  {\bibfnamefont {Philip}\ \bibnamefont {Kim}}, \bibinfo {author}
  {\bibfnamefont {Michael~L.}\ \bibnamefont {Steigerwald}}, \bibinfo {author}
  {\bibfnamefont {Jingjing}\ \bibnamefont {Yang}}, \bibinfo {author}
  {\bibfnamefont {Colin}\ \bibnamefont {Nuckolls}}, \ and\ \bibinfo {author}
  {\bibfnamefont {Xavier}\ \bibnamefont {Roy}},\ }\bibfield  {title} {\enquote
  {\bibinfo {title} {A few-layer covalent network of fullerenes},}\ }\href
  {\doibase 10.1038/s41586-022-05401-w} {\bibfield  {journal} {\bibinfo
  {journal} {Nature}\ }\textbf {\bibinfo {volume} {613}},\ \bibinfo {pages}
  {71--76} (\bibinfo {year} {2023})}\BibitemShut {NoStop}%
\bibitem [{\citenamefont {Wang}\ \emph {et~al.}(2023)\citenamefont {Wang},
  \citenamefont {Zhang}, \citenamefont {Wu}, \citenamefont {Chen},
  \citenamefont {Yang}, \citenamefont {Lu},\ and\ \citenamefont
  {Du}}]{Wang2023}%
  \BibitemOpen
  \bibfield  {author} {\bibinfo {author} {\bibfnamefont {Taotao}\ \bibnamefont
  {Wang}}, \bibinfo {author} {\bibfnamefont {Li}~\bibnamefont {Zhang}},
  \bibinfo {author} {\bibfnamefont {Jinbao}\ \bibnamefont {Wu}}, \bibinfo
  {author} {\bibfnamefont {Muqing}\ \bibnamefont {Chen}}, \bibinfo {author}
  {\bibfnamefont {Shangfeng}\ \bibnamefont {Yang}}, \bibinfo {author}
  {\bibfnamefont {Yalin}\ \bibnamefont {Lu}}, \ and\ \bibinfo {author}
  {\bibfnamefont {Pingwu}\ \bibnamefont {Du}},\ }\bibfield  {title} {\enquote
  {\bibinfo {title} {Few-layer fullerene network for photocatalytic pure water
  splitting into h$_2$ and h$_2$o$_2$},}\ }\href {\doibase
  10.1002/anie.202311352} {\bibfield  {journal} {\bibinfo  {journal} {Angew.
  Chem. Int. Ed.}\ }\textbf {\bibinfo {volume} {62}},\ \bibinfo {pages}
  {e202311352} (\bibinfo {year} {2023})}\BibitemShut {NoStop}%
\bibitem [{\citenamefont {Zhang}\ \emph
  {et~al.}(2025{\natexlab{a}})\citenamefont {Zhang}, \citenamefont {Xie},
  \citenamefont {Mei}, \citenamefont {Yu}, \citenamefont {Li}, \citenamefont
  {He}, \citenamefont {Fan}, \citenamefont {Zhang}, \citenamefont
  {Ricciardulli}, \citenamefont {Samor{\`i}}, \citenamefont {Li},\ and\
  \citenamefont {Yang}}]{Zhang2025}%
  \BibitemOpen
  \bibfield  {author} {\bibinfo {author} {\bibfnamefont {Yuxuan}\ \bibnamefont
  {Zhang}}, \bibinfo {author} {\bibfnamefont {Yifan}\ \bibnamefont {Xie}},
  \bibinfo {author} {\bibfnamefont {Hao}\ \bibnamefont {Mei}}, \bibinfo
  {author} {\bibfnamefont {Hui}\ \bibnamefont {Yu}}, \bibinfo {author}
  {\bibfnamefont {Minjuan}\ \bibnamefont {Li}}, \bibinfo {author}
  {\bibfnamefont {Zexiang}\ \bibnamefont {He}}, \bibinfo {author}
  {\bibfnamefont {Wentao}\ \bibnamefont {Fan}}, \bibinfo {author}
  {\bibfnamefont {Panpan}\ \bibnamefont {Zhang}}, \bibinfo {author}
  {\bibfnamefont {Antonio~Gaetano}\ \bibnamefont {Ricciardulli}}, \bibinfo
  {author} {\bibfnamefont {Paolo}\ \bibnamefont {Samor{\`i}}}, \bibinfo
  {author} {\bibfnamefont {Mengmeng}\ \bibnamefont {Li}}, \ and\ \bibinfo
  {author} {\bibfnamefont {Sheng}\ \bibnamefont {Yang}},\ }\bibfield  {title}
  {\enquote {\bibinfo {title} {Electrochemical synthesis of 2d polymeric
  fullerene for broadband photodetection},}\ }\href {\doibase
  10.1002/adma.202416741} {\bibfield  {journal} {\bibinfo  {journal} {Adv.
  Mater.}\ }\textbf {\bibinfo {volume} {37}},\ \bibinfo {pages} {2416741}
  (\bibinfo {year} {2025}{\natexlab{a}})}\BibitemShut {NoStop}%
\bibitem [{\citenamefont {Tomilin}\ \emph {et~al.}(2001)\citenamefont
  {Tomilin}, \citenamefont {Avramov}, \citenamefont {Varganov}, \citenamefont
  {Kuzubov},\ and\ \citenamefont {Ovchinnikov}}]{Tomilin2001}%
  \BibitemOpen
  \bibfield  {author} {\bibinfo {author} {\bibfnamefont {F.~N.}\ \bibnamefont
  {Tomilin}}, \bibinfo {author} {\bibfnamefont {P.~V.}\ \bibnamefont
  {Avramov}}, \bibinfo {author} {\bibfnamefont {S.~A.}\ \bibnamefont
  {Varganov}}, \bibinfo {author} {\bibfnamefont {A.~A.}\ \bibnamefont
  {Kuzubov}}, \ and\ \bibinfo {author} {\bibfnamefont {S.~G.}\ \bibnamefont
  {Ovchinnikov}},\ }\bibfield  {title} {\enquote {\bibinfo {title} {Possible
  scheme of synthesis-assembling of fullerenes},}\ }\href {\doibase
  10.1134/1.1371387} {\bibfield  {journal} {\bibinfo  {journal} {Physics of the
  Solid State}\ }\textbf {\bibinfo {volume} {43}},\ \bibinfo {pages} {973--981}
  (\bibinfo {year} {2001})}\BibitemShut {NoStop}%
\bibitem [{\citenamefont {Enyashin}\ and\ \citenamefont
  {Ivanovskii}(2008)}]{Enyashin2008}%
  \BibitemOpen
  \bibfield  {author} {\bibinfo {author} {\bibfnamefont {Andrey~N.}\
  \bibnamefont {Enyashin}}\ and\ \bibinfo {author} {\bibfnamefont
  {Alexander~L.}\ \bibnamefont {Ivanovskii}},\ }\bibfield  {title} {\enquote
  {\bibinfo {title} {{{Structural, electronic, cohesive, and elastic properties
  of diamondlike allotropes of crystalline ${\mathrm{C}}_{40}$}}},}\ }\href
  {\doibase 10.1103/PhysRevB.77.113402} {\bibfield  {journal} {\bibinfo
  {journal} {Phys. Rev. B}\ }\textbf {\bibinfo {volume} {77}},\ \bibinfo
  {pages} {113402} (\bibinfo {year} {2008})}\BibitemShut {NoStop}%
\bibitem [{\citenamefont {Kharlamov}\ \emph {et~al.}(2013)\citenamefont
  {Kharlamov}, \citenamefont {Kharlamova}, \citenamefont {Bondarenko},\ and\
  \citenamefont {Fomenko}}]{Kharlamov2013}%
  \BibitemOpen
  \bibfield  {author} {\bibinfo {author} {\bibfnamefont {Alexey}\ \bibnamefont
  {Kharlamov}}, \bibinfo {author} {\bibfnamefont {Ganna}\ \bibnamefont
  {Kharlamova}}, \bibinfo {author} {\bibfnamefont {Marina}\ \bibnamefont
  {Bondarenko}}, \ and\ \bibinfo {author} {\bibfnamefont {Veniamin}\
  \bibnamefont {Fomenko}},\ }\bibfield  {title} {\enquote {\bibinfo {title}
  {{{Joint Synthesis of Small Carbon Molecules (C$_3$-C$_{11}$),
  Quasi-Fullerenes (C$_{40}$, C$_{48}$, C$_{52}$) and their Hydrides}}},}\
  }\href {\doibase 10.12691/ces-1-3-1} {\bibfield  {journal} {\bibinfo
  {journal} {Chemical Engineering and Science}\ }\textbf {\bibinfo {volume}
  {1}},\ \bibinfo {pages} {32--40} (\bibinfo {year} {2013})}\BibitemShut
  {NoStop}%
\bibitem [{\citenamefont {Kittel}(1976)}]{Kittel1976}%
  \BibitemOpen
  \bibfield  {author} {\bibinfo {author} {\bibfnamefont {Charles}\ \bibnamefont
  {Kittel}},\ }\href@noop {} {\emph {\bibinfo {title} {Introduction to solid
  state physics}}}\ (\bibinfo  {publisher} {Wiley New York},\ \bibinfo {year}
  {1976})\BibitemShut {NoStop}%
\bibitem [{\citenamefont {{\v{S}}mejkal}\ \emph {et~al.}(2023)\citenamefont
  {{\v{S}}mejkal}, \citenamefont {Marmodoro}, \citenamefont {Ahn},
  \citenamefont {Gonz\'alez-Hern\'andez}, \citenamefont {Turek}, \citenamefont
  {Mankovsky}, \citenamefont {Ebert}, \citenamefont {D'Souza}, \citenamefont
  {{\v{S}}ipr}, \citenamefont {Sinova},\ and\ \citenamefont
  {Jungwirth}}]{Simejkal2023}%
  \BibitemOpen
  \bibfield  {author} {\bibinfo {author} {\bibfnamefont {Libor}\ \bibnamefont
  {{\v{S}}mejkal}}, \bibinfo {author} {\bibfnamefont {Alberto}\ \bibnamefont
  {Marmodoro}}, \bibinfo {author} {\bibfnamefont {Kyo-Hoon}\ \bibnamefont
  {Ahn}}, \bibinfo {author} {\bibfnamefont {Rafael}\ \bibnamefont
  {Gonz\'alez-Hern\'andez}}, \bibinfo {author} {\bibfnamefont {Ilja}\
  \bibnamefont {Turek}}, \bibinfo {author} {\bibfnamefont {Sergiy}\
  \bibnamefont {Mankovsky}}, \bibinfo {author} {\bibfnamefont {Hubert}\
  \bibnamefont {Ebert}}, \bibinfo {author} {\bibfnamefont {Sunil~W.}\
  \bibnamefont {D'Souza}}, \bibinfo {author} {\bibfnamefont {Ond{\v{r}}ej}\
  \bibnamefont {{\v{S}}ipr}}, \bibinfo {author} {\bibfnamefont {Jairo}\
  \bibnamefont {Sinova}}, \ and\ \bibinfo {author} {\bibfnamefont
  {Tom\'a{\v{s}}}\ \bibnamefont {Jungwirth}},\ }\bibfield  {title} {\enquote
  {\bibinfo {title} {{{Chiral Magnons in Altermagnetic
  ${\mathrm{RuO}}_{2}$}}},}\ }\href {\doibase 10.1103/PhysRevLett.131.256703}
  {\bibfield  {journal} {\bibinfo  {journal} {Phys. Rev. Lett.}\ }\textbf
  {\bibinfo {volume} {131}},\ \bibinfo {pages} {256703} (\bibinfo {year}
  {2023})}\BibitemShut {NoStop}%
\bibitem [{\citenamefont {Zhang}\ \emph
  {et~al.}(2025{\natexlab{b}})\citenamefont {Zhang}, \citenamefont {Ni},
  \citenamefont {Chen},\ and\ \citenamefont {Cao}}]{Zhang2025a}%
  \BibitemOpen
  \bibfield  {author} {\bibinfo {author} {\bibfnamefont {Yi-Fan}\ \bibnamefont
  {Zhang}}, \bibinfo {author} {\bibfnamefont {Xiao-Sheng}\ \bibnamefont {Ni}},
  \bibinfo {author} {\bibfnamefont {Ke}~\bibnamefont {Chen}}, \ and\ \bibinfo
  {author} {\bibfnamefont {Kun}\ \bibnamefont {Cao}},\ }\bibfield  {title}
  {\enquote {\bibinfo {title} {Chiral magnon splitting in altermagnetic crsb
  from first principles},}\ }\href {\doibase 10.1103/PhysRevB.111.174451}
  {\bibfield  {journal} {\bibinfo  {journal} {Phys. Rev. B}\ }\textbf {\bibinfo
  {volume} {111}},\ \bibinfo {pages} {174451} (\bibinfo {year}
  {2025}{\natexlab{b}})}\BibitemShut {NoStop}%
\bibitem [{\citenamefont {Tellez-Mora}\ \emph {et~al.}(2024)\citenamefont
  {Tellez-Mora}, \citenamefont {He}, \citenamefont {Bousquet}, \citenamefont
  {Wirtz},\ and\ \citenamefont {Romero}}]{Tellez-Mora2024}%
  \BibitemOpen
  \bibfield  {author} {\bibinfo {author} {\bibfnamefont {Andres}\ \bibnamefont
  {Tellez-Mora}}, \bibinfo {author} {\bibfnamefont {Xu}~\bibnamefont {He}},
  \bibinfo {author} {\bibfnamefont {Eric}\ \bibnamefont {Bousquet}}, \bibinfo
  {author} {\bibfnamefont {Ludger}\ \bibnamefont {Wirtz}}, \ and\ \bibinfo
  {author} {\bibfnamefont {Aldo~H.}\ \bibnamefont {Romero}},\ }\bibfield
  {title} {\enquote {\bibinfo {title} {Systematic determination of a material's
  magnetic ground state from first principles},}\ }\href {\doibase
  10.1038/s41524-024-01202-z} {\bibfield  {journal} {\bibinfo  {journal} {npj
  Computational Materials}\ }\textbf {\bibinfo {volume} {10}},\ \bibinfo
  {pages} {20} (\bibinfo {year} {2024})}\BibitemShut {NoStop}%
\bibitem [{\citenamefont {Klocke}\ \emph {et~al.}(2024)\citenamefont {Klocke},
  \citenamefont {Liu}, \citenamefont {Hal{\'a}sz},\ and\ \citenamefont
  {Alicea}}]{Klocke2024}%
  \BibitemOpen
  \bibfield  {author} {\bibinfo {author} {\bibfnamefont {Kai}\ \bibnamefont
  {Klocke}}, \bibinfo {author} {\bibfnamefont {Yue}\ \bibnamefont {Liu}},
  \bibinfo {author} {\bibfnamefont {G{\'a}bor~B}\ \bibnamefont {Hal{\'a}sz}}, \
  and\ \bibinfo {author} {\bibfnamefont {Jason}\ \bibnamefont {Alicea}},\
  }\bibfield  {title} {\enquote {\bibinfo {title} {Spin-liquid-based
  topological qubits},}\ }\href@noop {} {\bibfield  {journal} {\bibinfo
  {journal} {arXiv:2411.08093}\ } (\bibinfo {year} {2024})}\BibitemShut
  {NoStop}%
\bibitem [{\citenamefont {Zayed}\ \emph {et~al.}(2017)\citenamefont {Zayed},
  \citenamefont {R{\"u}egg}, \citenamefont {Larrea~J.}, \citenamefont
  {L{\"a}uchli}, \citenamefont {Panagopoulos}, \citenamefont {Saxena},
  \citenamefont {Ellerby}, \citenamefont {McMorrow}, \citenamefont
  {Str{\"a}ssle}, \citenamefont {Klotz}, \citenamefont {Hamel}, \citenamefont
  {Sadykov}, \citenamefont {Pomjakushin}, \citenamefont {Boehm}, \citenamefont
  {Jim{\'e}nez-Ruiz}, \citenamefont {Schneidewind}, \citenamefont
  {Pomjakushina}, \citenamefont {Stingaciu}, \citenamefont {Conder},\ and\
  \citenamefont {R{\/{o}}nnow}}]{Zayed2017}%
  \BibitemOpen
  \bibfield  {author} {\bibinfo {author} {\bibfnamefont {M.~E.}\ \bibnamefont
  {Zayed}}, \bibinfo {author} {\bibfnamefont {Ch.}\ \bibnamefont {R{\"u}egg}},
  \bibinfo {author} {\bibfnamefont {J.}~\bibnamefont {Larrea~J.}}, \bibinfo
  {author} {\bibfnamefont {A.~M.}\ \bibnamefont {L{\"a}uchli}}, \bibinfo
  {author} {\bibfnamefont {C.}~\bibnamefont {Panagopoulos}}, \bibinfo {author}
  {\bibfnamefont {S.~S.}\ \bibnamefont {Saxena}}, \bibinfo {author}
  {\bibfnamefont {M.}~\bibnamefont {Ellerby}}, \bibinfo {author} {\bibfnamefont
  {D.~F.}\ \bibnamefont {McMorrow}}, \bibinfo {author} {\bibfnamefont {Th.}\
  \bibnamefont {Str{\"a}ssle}}, \bibinfo {author} {\bibfnamefont
  {S.}~\bibnamefont {Klotz}}, \bibinfo {author} {\bibfnamefont
  {G.}~\bibnamefont {Hamel}}, \bibinfo {author} {\bibfnamefont {R.~A.}\
  \bibnamefont {Sadykov}}, \bibinfo {author} {\bibfnamefont {V.}~\bibnamefont
  {Pomjakushin}}, \bibinfo {author} {\bibfnamefont {M.}~\bibnamefont {Boehm}},
  \bibinfo {author} {\bibfnamefont {M.}~\bibnamefont {Jim{\'e}nez-Ruiz}},
  \bibinfo {author} {\bibfnamefont {A.}~\bibnamefont {Schneidewind}}, \bibinfo
  {author} {\bibfnamefont {E.}~\bibnamefont {Pomjakushina}}, \bibinfo {author}
  {\bibfnamefont {M.}~\bibnamefont {Stingaciu}}, \bibinfo {author}
  {\bibfnamefont {K.}~\bibnamefont {Conder}}, \ and\ \bibinfo {author}
  {\bibfnamefont {H.~M.}\ \bibnamefont {R{\/{o}}nnow}},\ }\bibfield  {title}
  {\enquote {\bibinfo {title} {4-spin plaquette singlet state in the
  shastry-sutherland compound srcu2(bo3)2},}\ }\href {\doibase
  10.1038/nphys4190} {\bibfield  {journal} {\bibinfo  {journal} {Nature
  Physics}\ }\textbf {\bibinfo {volume} {13}},\ \bibinfo {pages} {962--966}
  (\bibinfo {year} {2017})}\BibitemShut {NoStop}%
\bibitem [{\citenamefont {McClarty}\ \emph {et~al.}(2017)\citenamefont
  {McClarty}, \citenamefont {Kr{\"u}ger}, \citenamefont {Guidi}, \citenamefont
  {Parker}, \citenamefont {Refson}, \citenamefont {Parker}, \citenamefont
  {Prabhakaran},\ and\ \citenamefont {Coldea}}]{McClarty2017}%
  \BibitemOpen
  \bibfield  {author} {\bibinfo {author} {\bibfnamefont {P.~A.}\ \bibnamefont
  {McClarty}}, \bibinfo {author} {\bibfnamefont {F.}~\bibnamefont
  {Kr{\"u}ger}}, \bibinfo {author} {\bibfnamefont {T.}~\bibnamefont {Guidi}},
  \bibinfo {author} {\bibfnamefont {S.~F.}\ \bibnamefont {Parker}}, \bibinfo
  {author} {\bibfnamefont {K.}~\bibnamefont {Refson}}, \bibinfo {author}
  {\bibfnamefont {A.~W.}\ \bibnamefont {Parker}}, \bibinfo {author}
  {\bibfnamefont {D.}~\bibnamefont {Prabhakaran}}, \ and\ \bibinfo {author}
  {\bibfnamefont {R.}~\bibnamefont {Coldea}},\ }\bibfield  {title} {\enquote
  {\bibinfo {title} {{{Topological triplon modes and bound states in a
  Shastry-Sutherland magnet}}},}\ }\href {\doibase 10.1038/nphys4117}
  {\bibfield  {journal} {\bibinfo  {journal} {Nature Physics}\ }\textbf
  {\bibinfo {volume} {13}},\ \bibinfo {pages} {736--741} (\bibinfo {year}
  {2017})}\BibitemShut {NoStop}%
\bibitem [{\citenamefont {Wang}\ and\ \citenamefont
  {Batista}(2018)}]{Wang2018b}%
  \BibitemOpen
  \bibfield  {author} {\bibinfo {author} {\bibfnamefont {Zhentao}\ \bibnamefont
  {Wang}}\ and\ \bibinfo {author} {\bibfnamefont {Cristian~D.}\ \bibnamefont
  {Batista}},\ }\bibfield  {title} {\enquote {\bibinfo {title} {Dynamics and
  instabilities of the shastry-sutherland model},}\ }\href {\doibase
  10.1103/PhysRevLett.120.247201} {\bibfield  {journal} {\bibinfo  {journal}
  {Phys. Rev. Lett.}\ }\textbf {\bibinfo {volume} {120}},\ \bibinfo {pages}
  {247201} (\bibinfo {year} {2018})}\BibitemShut {NoStop}%
\bibitem [{\citenamefont {Wulferding}\ \emph {et~al.}(2021)\citenamefont
  {Wulferding}, \citenamefont {Choi}, \citenamefont {Lee}, \citenamefont
  {Prosnikov}, \citenamefont {Gallais}, \citenamefont {Lemmens}, \citenamefont
  {Zhong}, \citenamefont {Kageyama},\ and\ \citenamefont
  {Choi}}]{Wulferding2021}%
  \BibitemOpen
  \bibfield  {author} {\bibinfo {author} {\bibfnamefont {Dirk}\ \bibnamefont
  {Wulferding}}, \bibinfo {author} {\bibfnamefont {Youngsu}\ \bibnamefont
  {Choi}}, \bibinfo {author} {\bibfnamefont {Seungyeol}\ \bibnamefont {Lee}},
  \bibinfo {author} {\bibfnamefont {Mikhail~A.}\ \bibnamefont {Prosnikov}},
  \bibinfo {author} {\bibfnamefont {Yann}\ \bibnamefont {Gallais}}, \bibinfo
  {author} {\bibfnamefont {Peter}\ \bibnamefont {Lemmens}}, \bibinfo {author}
  {\bibfnamefont {Chengchao}\ \bibnamefont {Zhong}}, \bibinfo {author}
  {\bibfnamefont {Hiroshi}\ \bibnamefont {Kageyama}}, \ and\ \bibinfo {author}
  {\bibfnamefont {Kwang-Yong}\ \bibnamefont {Choi}},\ }\bibfield  {title}
  {\enquote {\bibinfo {title} {Thermally populated versus field-induced triplon
  bound states in the shastry-sutherland lattice srcu$_2$(bo$_3$)$_2$},}\
  }\href {\doibase 10.1038/s41535-021-00405-7} {\bibfield  {journal} {\bibinfo
  {journal} {npj Quantum Materials}\ }\textbf {\bibinfo {volume} {6}},\
  \bibinfo {pages} {102} (\bibinfo {year} {2021})}\BibitemShut {NoStop}%
\bibitem [{\citenamefont {Hohenberg}\ and\ \citenamefont
  {Kohn}(1964)}]{Hohenberg1964}%
  \BibitemOpen
  \bibfield  {author} {\bibinfo {author} {\bibfnamefont {P.}~\bibnamefont
  {Hohenberg}}\ and\ \bibinfo {author} {\bibfnamefont {W.}~\bibnamefont
  {Kohn}},\ }\bibfield  {title} {\enquote {\bibinfo {title} {Inhomogeneous
  electron gas},}\ }\href {\doibase 10.1103/PhysRev.136.B864} {\bibfield
  {journal} {\bibinfo  {journal} {Phys. Rev.}\ }\textbf {\bibinfo {volume}
  {136}},\ \bibinfo {pages} {B864--B871} (\bibinfo {year} {1964})}\BibitemShut
  {NoStop}%
\bibitem [{\citenamefont {Kohn}\ and\ \citenamefont {Sham}(1965)}]{Kohn1965}%
  \BibitemOpen
  \bibfield  {author} {\bibinfo {author} {\bibfnamefont {W.}~\bibnamefont
  {Kohn}}\ and\ \bibinfo {author} {\bibfnamefont {L.~J.}\ \bibnamefont
  {Sham}},\ }\bibfield  {title} {\enquote {\bibinfo {title} {Self-consistent
  equations including exchange and correlation effects},}\ }\href {\doibase
  10.1103/PhysRev.140.A1133} {\bibfield  {journal} {\bibinfo  {journal} {Phys.
  Rev.}\ }\textbf {\bibinfo {volume} {140}},\ \bibinfo {pages} {A1133--A1138}
  (\bibinfo {year} {1965})}\BibitemShut {NoStop}%
\bibitem [{\citenamefont {Kresse}\ and\ \citenamefont
  {Furthm\"uller}(1996{\natexlab{a}})}]{Kresse1996}%
  \BibitemOpen
  \bibfield  {author} {\bibinfo {author} {\bibfnamefont {G.}~\bibnamefont
  {Kresse}}\ and\ \bibinfo {author} {\bibfnamefont {J.}~\bibnamefont
  {Furthm\"uller}},\ }\bibfield  {title} {\enquote {\bibinfo {title}
  {{Efficient iterative schemes for \textit{ab initio} total-energy
  calculations using a plane-wave basis set}},}\ }\href {\doibase
  10.1103/PhysRevB.54.11169} {\bibfield  {journal} {\bibinfo  {journal} {Phys.
  Rev. B}\ }\textbf {\bibinfo {volume} {54}},\ \bibinfo {pages} {11169--11186}
  (\bibinfo {year} {1996}{\natexlab{a}})}\BibitemShut {NoStop}%
\bibitem [{\citenamefont {Kresse}\ and\ \citenamefont
  {Furthm\"uller}(1996{\natexlab{b}})}]{Kresse1996a}%
  \BibitemOpen
  \bibfield  {author} {\bibinfo {author} {\bibfnamefont {G.}~\bibnamefont
  {Kresse}}\ and\ \bibinfo {author} {\bibfnamefont {J.}~\bibnamefont
  {Furthm\"uller}},\ }\bibfield  {title} {\enquote {\bibinfo {title}
  {Efficiency of ab-initio total energy calculations for metals and
  semiconductors using a plane-wave basis set},}\ }\href {\doibase
  http://dx.doi.org/10.1016/0927-0256(96)00008-0} {\bibfield  {journal}
  {\bibinfo  {journal} {Computational Materials Science}\ }\textbf {\bibinfo
  {volume} {6}},\ \bibinfo {pages} {15 -- 50} (\bibinfo {year}
  {1996}{\natexlab{b}})}\BibitemShut {NoStop}%
\bibitem [{\citenamefont {Bl\"ochl}(1994)}]{Bloechl1994}%
  \BibitemOpen
  \bibfield  {author} {\bibinfo {author} {\bibfnamefont {P.~E.}\ \bibnamefont
  {Bl\"ochl}},\ }\bibfield  {title} {\enquote {\bibinfo {title} {Projector
  augmented-wave method},}\ }\href {\doibase 10.1103/PhysRevB.50.17953}
  {\bibfield  {journal} {\bibinfo  {journal} {Phys. Rev. B}\ }\textbf {\bibinfo
  {volume} {50}},\ \bibinfo {pages} {17953--17979} (\bibinfo {year}
  {1994})}\BibitemShut {NoStop}%
\bibitem [{\citenamefont {Kresse}\ and\ \citenamefont
  {Joubert}(1999)}]{Kresse1999}%
  \BibitemOpen
  \bibfield  {author} {\bibinfo {author} {\bibfnamefont {G.}~\bibnamefont
  {Kresse}}\ and\ \bibinfo {author} {\bibfnamefont {D.}~\bibnamefont
  {Joubert}},\ }\bibfield  {title} {\enquote {\bibinfo {title} {From ultrasoft
  pseudopotentials to the projector augmented-wave method},}\ }\href {\doibase
  10.1103/PhysRevB.59.1758} {\bibfield  {journal} {\bibinfo  {journal} {Phys.
  Rev. B}\ }\textbf {\bibinfo {volume} {59}},\ \bibinfo {pages} {1758--1775}
  (\bibinfo {year} {1999})}\BibitemShut {NoStop}%
\bibitem [{\citenamefont {Perdew}\ \emph {et~al.}(2008)\citenamefont {Perdew},
  \citenamefont {Ruzsinszky}, \citenamefont {Csonka}, \citenamefont {Vydrov},
  \citenamefont {Scuseria}, \citenamefont {Constantin}, \citenamefont {Zhou},\
  and\ \citenamefont {Burke}}]{Perdew2008}%
  \BibitemOpen
  \bibfield  {author} {\bibinfo {author} {\bibfnamefont {John~P.}\ \bibnamefont
  {Perdew}}, \bibinfo {author} {\bibfnamefont {Adrienn}\ \bibnamefont
  {Ruzsinszky}}, \bibinfo {author} {\bibfnamefont {G\'abor~I.}\ \bibnamefont
  {Csonka}}, \bibinfo {author} {\bibfnamefont {Oleg~A.}\ \bibnamefont
  {Vydrov}}, \bibinfo {author} {\bibfnamefont {Gustavo~E.}\ \bibnamefont
  {Scuseria}}, \bibinfo {author} {\bibfnamefont {Lucian~A.}\ \bibnamefont
  {Constantin}}, \bibinfo {author} {\bibfnamefont {Xiaolan}\ \bibnamefont
  {Zhou}}, \ and\ \bibinfo {author} {\bibfnamefont {Kieron}\ \bibnamefont
  {Burke}},\ }\bibfield  {title} {\enquote {\bibinfo {title} {{{Restoring the
  Density-Gradient Expansion for Exchange in Solids and Surfaces}}},}\ }\href
  {\doibase 10.1103/PhysRevLett.100.136406} {\bibfield  {journal} {\bibinfo
  {journal} {Phys. Rev. Lett.}\ }\textbf {\bibinfo {volume} {100}},\ \bibinfo
  {pages} {136406} (\bibinfo {year} {2008})}\BibitemShut {NoStop}%
\bibitem [{\citenamefont {Payne}\ \emph {et~al.}(1992)\citenamefont {Payne},
  \citenamefont {Teter}, \citenamefont {Allan}, \citenamefont {Arias},\ and\
  \citenamefont {Joannopoulos}}]{Payne1992}%
  \BibitemOpen
  \bibfield  {author} {\bibinfo {author} {\bibfnamefont {M.~C.}\ \bibnamefont
  {Payne}}, \bibinfo {author} {\bibfnamefont {M.~P.}\ \bibnamefont {Teter}},
  \bibinfo {author} {\bibfnamefont {D.~C.}\ \bibnamefont {Allan}}, \bibinfo
  {author} {\bibfnamefont {T.~A.}\ \bibnamefont {Arias}}, \ and\ \bibinfo
  {author} {\bibfnamefont {J.~D.}\ \bibnamefont {Joannopoulos}},\ }\bibfield
  {title} {\enquote {\bibinfo {title} {{{Iterative minimization techniques for
  \textit{ab initio} total-energy calculations: molecular dynamics and
  conjugate gradients}}},}\ }\href {\doibase 10.1103/RevModPhys.64.1045}
  {\bibfield  {journal} {\bibinfo  {journal} {Rev. Mod. Phys.}\ }\textbf
  {\bibinfo {volume} {64}},\ \bibinfo {pages} {1045--1097} (\bibinfo {year}
  {1992})}\BibitemShut {NoStop}%
\bibitem [{\citenamefont {Makov}\ and\ \citenamefont
  {Payne}(1995)}]{Makov1995}%
  \BibitemOpen
  \bibfield  {author} {\bibinfo {author} {\bibfnamefont {G.}~\bibnamefont
  {Makov}}\ and\ \bibinfo {author} {\bibfnamefont {M.~C.}\ \bibnamefont
  {Payne}},\ }\bibfield  {title} {\enquote {\bibinfo {title} {Periodic boundary
  conditions in ab initio calculations},}\ }\href {\doibase
  10.1103/PhysRevB.51.4014} {\bibfield  {journal} {\bibinfo  {journal} {Phys.
  Rev. B}\ }\textbf {\bibinfo {volume} {51}},\ \bibinfo {pages} {4014--4022}
  (\bibinfo {year} {1995})}\BibitemShut {NoStop}%
\bibitem [{\citenamefont {Marzari}\ and\ \citenamefont
  {Vanderbilt}(1997)}]{Marzari1997}%
  \BibitemOpen
  \bibfield  {author} {\bibinfo {author} {\bibfnamefont {Nicola}\ \bibnamefont
  {Marzari}}\ and\ \bibinfo {author} {\bibfnamefont {David}\ \bibnamefont
  {Vanderbilt}},\ }\bibfield  {title} {\enquote {\bibinfo {title} {{{Maximally
  localized generalized Wannier functions for composite energy bands}}},}\
  }\href {\doibase 10.1103/PhysRevB.56.12847} {\bibfield  {journal} {\bibinfo
  {journal} {Phys. Rev. B}\ }\textbf {\bibinfo {volume} {56}},\ \bibinfo
  {pages} {12847--12865} (\bibinfo {year} {1997})}\BibitemShut {NoStop}%
\bibitem [{\citenamefont {Souza}\ \emph {et~al.}(2001)\citenamefont {Souza},
  \citenamefont {Marzari},\ and\ \citenamefont {Vanderbilt}}]{Souza2001}%
  \BibitemOpen
  \bibfield  {author} {\bibinfo {author} {\bibfnamefont {Ivo}\ \bibnamefont
  {Souza}}, \bibinfo {author} {\bibfnamefont {Nicola}\ \bibnamefont {Marzari}},
  \ and\ \bibinfo {author} {\bibfnamefont {David}\ \bibnamefont {Vanderbilt}},\
  }\bibfield  {title} {\enquote {\bibinfo {title} {{{Maximally localized
  Wannier functions for entangled energy bands}}},}\ }\href {\doibase
  10.1103/PhysRevB.65.035109} {\bibfield  {journal} {\bibinfo  {journal} {Phys.
  Rev. B}\ }\textbf {\bibinfo {volume} {65}},\ \bibinfo {pages} {035109}
  (\bibinfo {year} {2001})}\BibitemShut {NoStop}%
\bibitem [{\citenamefont {Marzari}\ \emph {et~al.}(2012)\citenamefont
  {Marzari}, \citenamefont {Mostofi}, \citenamefont {Yates}, \citenamefont
  {Souza},\ and\ \citenamefont {Vanderbilt}}]{Marzari2012}%
  \BibitemOpen
  \bibfield  {author} {\bibinfo {author} {\bibfnamefont {Nicola}\ \bibnamefont
  {Marzari}}, \bibinfo {author} {\bibfnamefont {Arash~A.}\ \bibnamefont
  {Mostofi}}, \bibinfo {author} {\bibfnamefont {Jonathan~R.}\ \bibnamefont
  {Yates}}, \bibinfo {author} {\bibfnamefont {Ivo}\ \bibnamefont {Souza}}, \
  and\ \bibinfo {author} {\bibfnamefont {David}\ \bibnamefont {Vanderbilt}},\
  }\bibfield  {title} {\enquote {\bibinfo {title} {{{Maximally localized
  Wannier functions: Theory and applications}}},}\ }\href {\doibase
  10.1103/RevModPhys.84.1419} {\bibfield  {journal} {\bibinfo  {journal} {Rev.
  Mod. Phys.}\ }\textbf {\bibinfo {volume} {84}},\ \bibinfo {pages}
  {1419--1475} (\bibinfo {year} {2012})}\BibitemShut {NoStop}%
\bibitem [{\citenamefont {Mostofi}\ \emph {et~al.}(2008)\citenamefont
  {Mostofi}, \citenamefont {Yates}, \citenamefont {Lee}, \citenamefont {Souza},
  \citenamefont {Vanderbilt},\ and\ \citenamefont {Marzari}}]{Mostofi2008}%
  \BibitemOpen
  \bibfield  {author} {\bibinfo {author} {\bibfnamefont {Arash~A.}\
  \bibnamefont {Mostofi}}, \bibinfo {author} {\bibfnamefont {Jonathan~R.}\
  \bibnamefont {Yates}}, \bibinfo {author} {\bibfnamefont {Young-Su}\
  \bibnamefont {Lee}}, \bibinfo {author} {\bibfnamefont {Ivo}\ \bibnamefont
  {Souza}}, \bibinfo {author} {\bibfnamefont {David}\ \bibnamefont
  {Vanderbilt}}, \ and\ \bibinfo {author} {\bibfnamefont {Nicola}\ \bibnamefont
  {Marzari}},\ }\bibfield  {title} {\enquote {\bibinfo {title} {{{Wannier90: A
  tool for obtaining maximally-localised Wannier functions}}},}\ }\href
  {\doibase 10.1016/j.cpc.2007.11.016} {\bibfield  {journal} {\bibinfo
  {journal} {Computer Physics Communications}\ }\textbf {\bibinfo {volume}
  {178}},\ \bibinfo {pages} {685--699} (\bibinfo {year} {2008})}\BibitemShut
  {NoStop}%
\bibitem [{\citenamefont {Mostofi}\ \emph {et~al.}(2014)\citenamefont
  {Mostofi}, \citenamefont {Yates}, \citenamefont {Pizzi}, \citenamefont {Lee},
  \citenamefont {Souza}, \citenamefont {Vanderbilt},\ and\ \citenamefont
  {Marzari}}]{Mostofi2014}%
  \BibitemOpen
  \bibfield  {author} {\bibinfo {author} {\bibfnamefont {Arash~A.}\
  \bibnamefont {Mostofi}}, \bibinfo {author} {\bibfnamefont {Jonathan~R.}\
  \bibnamefont {Yates}}, \bibinfo {author} {\bibfnamefont {Giovanni}\
  \bibnamefont {Pizzi}}, \bibinfo {author} {\bibfnamefont {Young-Su}\
  \bibnamefont {Lee}}, \bibinfo {author} {\bibfnamefont {Ivo}\ \bibnamefont
  {Souza}}, \bibinfo {author} {\bibfnamefont {David}\ \bibnamefont
  {Vanderbilt}}, \ and\ \bibinfo {author} {\bibfnamefont {Nicola}\ \bibnamefont
  {Marzari}},\ }\bibfield  {title} {\enquote {\bibinfo {title} {{{An updated
  version of Wannier90: A tool for obtaining maximally-localised Wannier
  functions}}},}\ }\href {\doibase 10.1016/j.cpc.2014.05.003} {\bibfield
  {journal} {\bibinfo  {journal} {Computer Physics Communications}\ }\textbf
  {\bibinfo {volume} {185}},\ \bibinfo {pages} {2309--2310} (\bibinfo {year}
  {2014})}\BibitemShut {NoStop}%
\bibitem [{\citenamefont {Pizzi}\ \emph {et~al.}(2020)\citenamefont {Pizzi},
  \citenamefont {Vitale}, \citenamefont {Arita}, \citenamefont {Bl{\"u}gel},
  \citenamefont {Freimuth}, \citenamefont {G{\'e}ranton}, \citenamefont
  {Gibertini}, \citenamefont {Gresch}, \citenamefont {Johnson}, \citenamefont
  {Koretsune}, \citenamefont {Iba{\~n}ez-Azpiroz}, \citenamefont {Lee},
  \citenamefont {Lihm}, \citenamefont {Marchand}, \citenamefont {Marrazzo},
  \citenamefont {Mokrousov}, \citenamefont {Mustafa}, \citenamefont {Nohara},
  \citenamefont {Nomura}, \citenamefont {Paulatto}, \citenamefont {Ponc{\'e}},
  \citenamefont {Ponweiser}, \citenamefont {Qiao}, \citenamefont {Th{\"o}le},
  \citenamefont {Tsirkin}, \citenamefont {Wierzbowska}, \citenamefont
  {Marzari}, \citenamefont {Vanderbilt}, \citenamefont {Souza}, \citenamefont
  {Mostofi},\ and\ \citenamefont {Yates}}]{Pizzi2020}%
  \BibitemOpen
  \bibfield  {author} {\bibinfo {author} {\bibfnamefont {Giovanni}\
  \bibnamefont {Pizzi}}, \bibinfo {author} {\bibfnamefont {Valerio}\
  \bibnamefont {Vitale}}, \bibinfo {author} {\bibfnamefont {Ryotaro}\
  \bibnamefont {Arita}}, \bibinfo {author} {\bibfnamefont {Stefan}\
  \bibnamefont {Bl{\"u}gel}}, \bibinfo {author} {\bibfnamefont {Frank}\
  \bibnamefont {Freimuth}}, \bibinfo {author} {\bibfnamefont {Guillaume}\
  \bibnamefont {G{\'e}ranton}}, \bibinfo {author} {\bibfnamefont {Marco}\
  \bibnamefont {Gibertini}}, \bibinfo {author} {\bibfnamefont {Dominik}\
  \bibnamefont {Gresch}}, \bibinfo {author} {\bibfnamefont {Charles}\
  \bibnamefont {Johnson}}, \bibinfo {author} {\bibfnamefont {Takashi}\
  \bibnamefont {Koretsune}}, \bibinfo {author} {\bibfnamefont {Julen}\
  \bibnamefont {Iba{\~n}ez-Azpiroz}}, \bibinfo {author} {\bibfnamefont
  {Hyungjun}\ \bibnamefont {Lee}}, \bibinfo {author} {\bibfnamefont {Jae-Mo}\
  \bibnamefont {Lihm}}, \bibinfo {author} {\bibfnamefont {Daniel}\ \bibnamefont
  {Marchand}}, \bibinfo {author} {\bibfnamefont {Antimo}\ \bibnamefont
  {Marrazzo}}, \bibinfo {author} {\bibfnamefont {Yuriy}\ \bibnamefont
  {Mokrousov}}, \bibinfo {author} {\bibfnamefont {Jamal~I}\ \bibnamefont
  {Mustafa}}, \bibinfo {author} {\bibfnamefont {Yoshiro}\ \bibnamefont
  {Nohara}}, \bibinfo {author} {\bibfnamefont {Yusuke}\ \bibnamefont {Nomura}},
  \bibinfo {author} {\bibfnamefont {Lorenzo}\ \bibnamefont {Paulatto}},
  \bibinfo {author} {\bibfnamefont {Samuel}\ \bibnamefont {Ponc{\'e}}},
  \bibinfo {author} {\bibfnamefont {Thomas}\ \bibnamefont {Ponweiser}},
  \bibinfo {author} {\bibfnamefont {Junfeng}\ \bibnamefont {Qiao}}, \bibinfo
  {author} {\bibfnamefont {Florian}\ \bibnamefont {Th{\"o}le}}, \bibinfo
  {author} {\bibfnamefont {Stepan~S}\ \bibnamefont {Tsirkin}}, \bibinfo
  {author} {\bibfnamefont {Malgorzata}\ \bibnamefont {Wierzbowska}}, \bibinfo
  {author} {\bibfnamefont {Nicola}\ \bibnamefont {Marzari}}, \bibinfo {author}
  {\bibfnamefont {David}\ \bibnamefont {Vanderbilt}}, \bibinfo {author}
  {\bibfnamefont {Ivo}\ \bibnamefont {Souza}}, \bibinfo {author} {\bibfnamefont
  {Arash~A}\ \bibnamefont {Mostofi}}, \ and\ \bibinfo {author} {\bibfnamefont
  {Jonathan~R}\ \bibnamefont {Yates}},\ }\bibfield  {title} {\enquote {\bibinfo
  {title} {{{Wannier90 as a community code: new features and applications}}},}\
  }\href {\doibase 10.1088/1361-648X/ab51ff} {\bibfield  {journal} {\bibinfo
  {journal} {J. Phys.: Condens. Matter}\ }\textbf {\bibinfo {volume} {32}},\
  \bibinfo {pages} {165902} (\bibinfo {year} {2020})}\BibitemShut {NoStop}%
\bibitem [{\citenamefont {Liechtenstein}\ \emph {et~al.}(1987)\citenamefont
  {Liechtenstein}, \citenamefont {Katsnelson}, \citenamefont {Antropov},\ and\
  \citenamefont {Gubanov}}]{Liechtenstein1987}%
  \BibitemOpen
  \bibfield  {author} {\bibinfo {author} {\bibfnamefont {A.I.}\ \bibnamefont
  {Liechtenstein}}, \bibinfo {author} {\bibfnamefont {M.I.}\ \bibnamefont
  {Katsnelson}}, \bibinfo {author} {\bibfnamefont {V.P.}\ \bibnamefont
  {Antropov}}, \ and\ \bibinfo {author} {\bibfnamefont {V.A.}\ \bibnamefont
  {Gubanov}},\ }\bibfield  {title} {\enquote {\bibinfo {title} {Local spin
  density functional approach to the theory of exchange interactions in
  ferromagnetic metals and alloys},}\ }\href {\doibase
  10.1016/0304-8853(87)90721-9} {\bibfield  {journal} {\bibinfo  {journal}
  {Journal of Magnetism and Magnetic Materials}\ }\textbf {\bibinfo {volume}
  {67}},\ \bibinfo {pages} {65--74} (\bibinfo {year} {1987})}\BibitemShut
  {NoStop}%
\bibitem [{\citenamefont {Korotin}\ \emph {et~al.}(2015)\citenamefont
  {Korotin}, \citenamefont {Mazurenko}, \citenamefont {Anisimov},\ and\
  \citenamefont {Streltsov}}]{Korotin2015}%
  \BibitemOpen
  \bibfield  {author} {\bibinfo {author} {\bibfnamefont {Dm.~M.}\ \bibnamefont
  {Korotin}}, \bibinfo {author} {\bibfnamefont {V.~V.}\ \bibnamefont
  {Mazurenko}}, \bibinfo {author} {\bibfnamefont {V.~I.}\ \bibnamefont
  {Anisimov}}, \ and\ \bibinfo {author} {\bibfnamefont {S.~V.}\ \bibnamefont
  {Streltsov}},\ }\bibfield  {title} {\enquote {\bibinfo {title} {Calculation
  of exchange constants of the heisenberg model in plane-wave-based methods
  using the green's function approach},}\ }\href {\doibase
  10.1103/PhysRevB.91.224405} {\bibfield  {journal} {\bibinfo  {journal} {Phys.
  Rev. B}\ }\textbf {\bibinfo {volume} {91}},\ \bibinfo {pages} {224405}
  (\bibinfo {year} {2015})}\BibitemShut {NoStop}%
\bibitem [{\citenamefont {He}\ \emph {et~al.}(2021)\citenamefont {He},
  \citenamefont {Helbig}, \citenamefont {Verstraete},\ and\ \citenamefont
  {Bousquet}}]{He2021}%
  \BibitemOpen
  \bibfield  {author} {\bibinfo {author} {\bibfnamefont {Xu}~\bibnamefont
  {He}}, \bibinfo {author} {\bibfnamefont {Nicole}\ \bibnamefont {Helbig}},
  \bibinfo {author} {\bibfnamefont {Matthieu~J.}\ \bibnamefont {Verstraete}}, \
  and\ \bibinfo {author} {\bibfnamefont {Eric}\ \bibnamefont {Bousquet}},\
  }\bibfield  {title} {\enquote {\bibinfo {title} {Tb2j: A python package for
  computing magnetic interaction parameters},}\ }\href {\doibase
  10.1016/j.cpc.2021.107938} {\bibfield  {journal} {\bibinfo  {journal}
  {Computer Physics Communications}\ }\textbf {\bibinfo {volume} {264}},\
  \bibinfo {pages} {107938} (\bibinfo {year} {2021})}\BibitemShut {NoStop}%
\bibitem [{\citenamefont {Holstein}\ and\ \citenamefont
  {Primakoff}(1940)}]{Holstein1940}%
  \BibitemOpen
  \bibfield  {author} {\bibinfo {author} {\bibfnamefont {T.}~\bibnamefont
  {Holstein}}\ and\ \bibinfo {author} {\bibfnamefont {H.}~\bibnamefont
  {Primakoff}},\ }\bibfield  {title} {\enquote {\bibinfo {title} {Field
  dependence of the intrinsic domain magnetization of a ferromagnet},}\ }\href
  {\doibase 10.1103/PhysRev.58.1098} {\bibfield  {journal} {\bibinfo  {journal}
  {Phys. Rev.}\ }\textbf {\bibinfo {volume} {58}},\ \bibinfo {pages}
  {1098--1113} (\bibinfo {year} {1940})}\BibitemShut {NoStop}%
\bibitem [{\citenamefont {Colpa}(1978)}]{Colpa1978}%
  \BibitemOpen
  \bibfield  {author} {\bibinfo {author} {\bibfnamefont {J.H.P.}\ \bibnamefont
  {Colpa}},\ }\bibfield  {title} {\enquote {\bibinfo {title} {Diagonalization
  of the quadratic boson hamiltonian},}\ }\href {\doibase
  10.1016/0378-4371(78)90160-7} {\bibfield  {journal} {\bibinfo  {journal}
  {Physica A: Statistical Mechanics and its Applications}\ }\textbf {\bibinfo
  {volume} {93}},\ \bibinfo {pages} {327--353} (\bibinfo {year}
  {1978})}\BibitemShut {NoStop}%
\bibitem [{\citenamefont {Rybakov}\ \emph {et~al.}(2024)\citenamefont
  {Rybakov}, \citenamefont {Boix-Constant}, \citenamefont {Alba~Venero},
  \citenamefont {van~der Zant}, \citenamefont {Ma{\~n}as-Valero},\ and\
  \citenamefont {Coronado}}]{Rybakov2024}%
  \BibitemOpen
  \bibfield  {author} {\bibinfo {author} {\bibfnamefont {Andrey}\ \bibnamefont
  {Rybakov}}, \bibinfo {author} {\bibfnamefont {Carla}\ \bibnamefont
  {Boix-Constant}}, \bibinfo {author} {\bibfnamefont {Diego}\ \bibnamefont
  {Alba~Venero}}, \bibinfo {author} {\bibfnamefont {Herre S.~J.}\ \bibnamefont
  {van~der Zant}}, \bibinfo {author} {\bibfnamefont {Samuel}\ \bibnamefont
  {Ma{\~n}as-Valero}}, \ and\ \bibinfo {author} {\bibfnamefont {Eugenio}\
  \bibnamefont {Coronado}},\ }\bibfield  {title} {\enquote {\bibinfo {title}
  {Probing short-range correlations in the van der waals magnet crsbr by
  small-angle neutron scattering},}\ }\href {\doibase 10.1002/smsc.202400244}
  {\bibfield  {journal} {\bibinfo  {journal} {Small Sci.}\ }\textbf {\bibinfo
  {volume} {4}},\ \bibinfo {pages} {2400244--} (\bibinfo {year}
  {2024})}\BibitemShut {NoStop}%
\bibitem [{\citenamefont {Boix-Constant}\ \emph {et~al.}(2025)\citenamefont
  {Boix-Constant}, \citenamefont {Rybakov}, \citenamefont {Miranda-P{\'e}rez},
  \citenamefont {Mart{\'i}nez-Carracedo}, \citenamefont {Ferrer}, \citenamefont
  {Ma{\~n}as-Valero},\ and\ \citenamefont {Coronado}}]{Boix-Constant2025}%
  \BibitemOpen
  \bibfield  {author} {\bibinfo {author} {\bibfnamefont {Carla}\ \bibnamefont
  {Boix-Constant}}, \bibinfo {author} {\bibfnamefont {Andrey}\ \bibnamefont
  {Rybakov}}, \bibinfo {author} {\bibfnamefont {Clara}\ \bibnamefont
  {Miranda-P{\'e}rez}}, \bibinfo {author} {\bibfnamefont {Gabriel}\
  \bibnamefont {Mart{\'i}nez-Carracedo}}, \bibinfo {author} {\bibfnamefont
  {Jaime}\ \bibnamefont {Ferrer}}, \bibinfo {author} {\bibfnamefont {Samuel}\
  \bibnamefont {Ma{\~n}as-Valero}}, \ and\ \bibinfo {author} {\bibfnamefont
  {Eugenio}\ \bibnamefont {Coronado}},\ }\bibfield  {title} {\enquote {\bibinfo
  {title} {Programmable magnetic hysteresis in orthogonally-twisted 2d crsbr
  magnets via stacking engineering.}}\ }\href {\doibase 10.1002/adma.202415774}
  {\bibfield  {journal} {\bibinfo  {journal} {Adv. Mater.}\ }\textbf {\bibinfo
  {volume} {37}},\ \bibinfo {pages} {2415774--} (\bibinfo {year}
  {2025})}\BibitemShut {NoStop}%
\bibitem [{\citenamefont {Yuan}\ \emph {et~al.}(2025)\citenamefont {Yuan},
  \citenamefont {Jankowski}, \citenamefont {Shen},\ and\ \citenamefont
  {Slager}}]{Yuan2025}%
  \BibitemOpen
  \bibfield  {author} {\bibinfo {author} {\bibfnamefont {Rundong}\ \bibnamefont
  {Yuan}}, \bibinfo {author} {\bibfnamefont {Wojciech~J}\ \bibnamefont
  {Jankowski}}, \bibinfo {author} {\bibfnamefont {Ka}~\bibnamefont {Shen}}, \
  and\ \bibinfo {author} {\bibfnamefont {Robert-Jan}\ \bibnamefont {Slager}},\
  }\bibfield  {title} {\enquote {\bibinfo {title} {Quantum geometry of
  altermagnetic magnons probed by light},}\ }\href@noop {} {\bibfield
  {journal} {\bibinfo  {journal} {arXiv:2508.02781}\ } (\bibinfo {year}
  {2025})}\BibitemShut {NoStop}%
\bibitem [{\citenamefont {Adamo}\ and\ \citenamefont
  {Barone}(1999)}]{Adamo1999}%
  \BibitemOpen
  \bibfield  {author} {\bibinfo {author} {\bibfnamefont {Carlo}\ \bibnamefont
  {Adamo}}\ and\ \bibinfo {author} {\bibfnamefont {Vincenzo}\ \bibnamefont
  {Barone}},\ }\bibfield  {title} {\enquote {\bibinfo {title} {Toward reliable
  density functional methods without adjustable parameters: The pbe0 model},}\
  }\href {\doibase 10.1063/1.478522} {\bibfield  {journal} {\bibinfo  {journal}
  {J. Chem. Phys.}\ }\textbf {\bibinfo {volume} {110}},\ \bibinfo {pages}
  {6158--6170} (\bibinfo {year} {1999})}\BibitemShut {NoStop}%
\end{thebibliography}%

\end{document}